\documentclass{aa}  

\usepackage{graphicx}
\usepackage{txfonts}
\usepackage{graphicx} 
\graphicspath{{./figures/}} 
\usepackage[pdfpagelabels=false]{hyperref}
\hypersetup{colorlinks=true,linkcolor=blue,citecolor=blue,filecolor=blue,urlcolor=blue,}
\usepackage{multirow}
\usepackage[normalem]{ulem}

\newcommand{\centered}[1]{\begin{tabular}{l}#1\end{tabular}}
\defcitealias{gadotti+09}{G09}

\begin{document}

   \title{Revisiting the role of bars in AGN fuelling with propensity score sample matching}
   \titlerunning{Revisiting the role of bars in AGN fueling}

   \author{Luiz A. Silva-Lima\inst{1}
          \and
          Lucimara P. Martins\inst{1}
          \and
          Paula R. T. Coelho\inst{2}
          \and
          Dimitri A. Gadotti\inst{3}
          }

   \institute{NAT -- Universidade Cidade de São Paulo/Universidade Cruzeiro do Sul,  Rua Galv\~ao Bueno 868, S\~ao Paulo-SP, 01506-000, Brazil\\
              \email{luiz.sl@outlook.com}
        \and
            IAG -- Universidade de S\~ao Paulo, Rua do Mat\~ao, 1226, 05508-090, S\~ao Paulo-SP, Brazil
        \and
            European Southern Observatory, Karl-Schwarzschild-Str. 2, D-85748 Garching bei München, Germany
             }

   \date{Received xxxxxx xx, xxxx; accepted xxxxxxx xx, xxxx}

 
  \abstract
  {The high luminosity displayed by an active galactic nucleus (AGN) requires that gas be transported to the centre of the galaxy by some mechanism. Bar-driven processes are often pointed out in this context and a number of studies have addressed the bar--AGN connection, but with conflicting results. Some of the inconsistencies can be explained by the different spatial- and timescales involved in bar-driven gas inflows, accretion by the central black hole, and AGN emission. However, the discrepant results could also be due to sample biases, because both the AGN activity determination and the bar detection are influenced by the method employed. We revisit the bar--AGN connection in a sample of galaxies from SDSS, looking for evidence of the influence of bars on AGN activity. We determine AGN activity by emission line diagnostics and the properties of the bar were previously estimated with \texttt{BUDDA}, which performs 2D bulge--bar--disk decomposition. Before comparing active and inactive galaxies, we made a careful selection of the sample to minimise selection biases. We created control samples by matching them with the AGN sample using propensity score matching. This technique offers an analytical approach for creating control samples given some object parameters. We find that AGN are preferentially found in barred galaxies and that the accretion rate is higher in barred galaxies, but only when different M--$\sigma$ relations are used to estimate the black hole mass M$_\bullet$ in barred and unbarred galaxies (from the central velocity dispersion $\sigma$). On the other hand, we find no correlation between activity level and bar strength. Altogether, our results strengthen theoretical predictions that the bar is an important mechanism in disc galaxies, creating a gas reservoir to feed AGN, but they also indicate that other mechanisms can play a major role, particularly at scales\,$\lesssim$\,100\,pc.}

   \keywords{galaxies: active -- 
                galaxies: nuclei -- 
                galaxies: evolution -- 
                galaxies: structure
               }

   \maketitle
%

\section{Introduction}

Active galaxies present much higher luminosities than inactive galaxies over most of the electromagnetic spectrum, and the compact region at their centre is called an active galactic nucleus (AGN). Most of the energy output in AGN is a non-thermal (not stellar) type of emission. Nowadays, this phenomenon is understood as the effect of a supermassive black hole (SMBH) surrounded by a gaseous accretion disc. Such discs form because the material being accreted by the black hole has residual angular momentum. The disc can have differential rotation, creating friction between the rotating gas layers, which rises the temperature to millions of degrees and ultimately leads to a very high luminosity. Despite being present in most galactic centres \citep{kormendy+95}, only a fraction of SMBHs consume gas and stars in large quantities. For example, only 10\% of local galaxies are active. However, they were much more numerous in the past and possibly had a fundamental role in shaping the galaxies we see today \citep{ho+97}.

The luminosity emitted by an AGN depends on the accretion rate and the mass--energy efficiency, which is usually around 10\% \citep{peterson97, combes01, bian_2003}. Based on this, one can estimate the mass necessary to sustain this luminosity: for an AGN with a luminosity of L~$\approx$~10$^{44}$ erg s$^{-1}$, typical of a Seyfert galaxy, the total mass that has to be transported to the nuclear region varies between 10$^5$ and 10$^6$~M$_\odot$, considering a lifetime of 10$^7$ - 10$^8$ years for an AGN \citep{martini04, wada04, merloni04}. One of the major challenges in astronomy today is to understand how this large quantity of material is transported to the galactic centre, entering the black hole gravitational influence radius. To be moved from galactic scale orbits (tens of kiloparsecs (kpc)) to parcec(pc)-scale orbits, the gas has to lose angular momentum. 

Galaxy dynamics is very complex. Different dynamic instability mechanisms can have a role in the angular momentum removal \citep{combes01}, such as type m=1 instabilities (spirals or tide interaction with nearby galaxies) and type m=2 instabilities (bars). Dynamical mechanisms that could cause these perturbations have been investigated for a possible connection with AGN activity \citep{Regan+99, Laine+02, cisternas+13, depropris2014, satyapal2014, cheung+15, galloway+15, alonso+18, ellison2019}.

In particular, on kpc scales, the bar non-axisymmetric potential has an important role in the secular evolution of galaxies. Simulations suggest that torques produced by the stellar bar potential perturb the interstellar medium (ISM), transporting gas and dust through the bar, where this material suffers shocks and compression \citep{athanassoula92}. Observational evidence also points to the bars as a mechanism that transports gas to the central regions, increasing the young stellar population in the region of the bulge \citep{seth2005, coelho+11, ellison+11, oh+12, lin2020}, altering the gas content and N/O gas abundances \citep{florido+15}, or creating a correlation between the star formation rate (SFR) close to the nucleus and some bar properties \citep{martin95}. 
The influence of the bar through its ability to produce torque in the ISM gas is well accepted. However, for an intermediate scale of hundreds of pc, the gas finds a zone full of resonances as an obstacle to the direct feeding of the AGN. In this region, the gas is confined to rings near inner Lindblad resonances \citep{athanassoula92, piner1995, combes01, binney+09, sormani+2018, audibert2019}.

Recent studies investigated the bar--AGN connection, with controversial results. From a non-exhaustive review of recent results, we find that
\citet{oh+12} analysed the effect of the presence of bars on both star formation and AGN activity in the local Universe. They used a Sloan Digital Sky Survey (SDSS) sample of more than 6000 galaxies, using the BPT diagram \citep{baldwin1981} for activity detection and visual inspection for the presence of bars. \citet{oh+12} found an excess of AGN in barred galaxies. Their results suggest that the bar has an influence on the AGN and this effect is more intense on bluer galaxies and with less massive black holes. \citet{alonso13}, using a SDSS sample, find evidence that bars play an important role in AGN activity and host galaxy properties. In addition, the results reported by \citet{alonso14} point to a relationship between the environment and the efficiency of the bars in transporting gas towards the centre, with a higher efficiency encountered in denser environments. This suggests that analysis of a possible bar--AGN relationship should also take into account the type of environment of the galaxies in the sample. Another point that stresses the need to consider the environment in this study is that mergers and other types of interaction are also suggested as AGN feeding mechanisms. \citet{ellison+2016} report observational results that indicate that secular and interacting processes lead to distinct manifestations of activity. While secular processes predominantly lead to more moderate accretion rates and are not accompanied by an increase in SFR, interaction processes lead to an increase in SFR, with more powerful and possibly obscured AGN. \citet{ellison2019} also report observational results that support the scenario where interactions promote gas funnelling to the SMBH feed, with an AGN sample exhibiting more frequent disturbance signals than a sample of inactive galaxies. However, these authors emphasise that despite the evidence in favour of a merger--AGN connection, interactions and mergers should not be the predominant mechanisms in triggering the feeding of optically identified AGN in the local Universe.

\citet{alonso+18} also used a SDSS sample to study the influence of strong bars on AGN, while also comparing the effects of interactions on activity. These authors found that both bars and interactions increase the AGN luminosity and accretion rate, but bars have a greater efficiency in the process. Using a much larger sample, which makes use of the citizen science project Galaxy Zoo, \citet{galloway+15} also find that there is an excess of AGN in barred galaxies. However, \citet{cheung+15}, who also based their study on Galaxy Zoo, but with a sample selected beyond the local Universe (0.2 $\leq$ z $\leq$ 1.0), find no excess of AGN in strongly barred galaxies. \citet{cisternas+13}, using a sample of X-ray-luminosity-selected AGN, find similar results: the fraction of AGN in barred galaxies seems to be higher, but there is no correlation between the AGN strength (measured by the X-ray luminosity) and the presence of a bar. \citet{goulding+17}, making use of a large sample of galaxies from the SDSS Galaxy Zoo project 2 and the X-ray stacking analysis technique to measure BH accretion rates for more than 50,000 galaxies with and without strong bars, again find no evidence that large-scale bars influence the average growth of BHs in nearby galaxies.

In summary, we find that while \citet{Regan+99}, \citet{cisternas+13}, \citet{cheung+15}, and \citet{goulding+17} do not find significant differences that indicate bars as the main mechanism for AGN feeding, \citet{Laine+02}, \citet{coelho+11}, \citet{oh+12}, \citet{alonso13, alonso14}, \citet{galloway+15}, and \citet{alonso+18} show that there seems to be a connection between bars and AGN. These conflicting results might be explained by the difficulties in the sample selection and characterisation, and intrinsic complications predicted theoretically. From an observational point of view, bar and activity classification can be determined by different methodologies, and these do not always agree. For the activity classification, one of the most used criteria is the BTP diagram \citep{baldwin1981}, together with the divisory lines from \citet{Kauffmann2003}, \citet{Kewley2001}, and \citet{Schawinski2007}. Other methods can also be used, like the full width of half maximum (FWHM) of the H$\alpha$ line \citep{hao+09}, X-ray luminosity \citep[e.g.][]{goulding+17}, or colours \citep{stern+12}. 
Bar identification is also prone to uncertainties. 
Identification can be made through visual inspection \citep[e.g.][]{oh+12}, model fitting \citep[e.g.][]{menendez-delmestre2007, gadotti+09}, or large citizen science projects like Galaxy Zoo \citep[e.g.][]{galloway+15}. 
In a recent study, \cite{lee2019}, studying a volume-limited sample consisting of 1698 spiral galaxies brighter than M$_r$ = -15.2 with z $<$ 0.01 from SDSS/DR7, 
demonstrate that the fraction of bars is dependent on the detection method, finding bar fractions of $63\%$ when using visual inspection, $48\%$ when fitting ellipses, and $36\%$ when using Fourier analysis.

Also, the cosmological distribution of bars needs to be taken with care. 
After the pioneering work by \citet{abraham+99}, many authors sought to study the evolution of the barred galaxy fraction with redshift \citep{jogee+04, sheth+08, cameron+10, sheth+12, melvin+14, simmons+14}. While the results between the different studies are not exactly compatible with each other, there seems to be a consensus that there is a decline in the numerical density of bars from the local Universe up to z $\approx$ 0.8. 

If on one hand, observational constraints such as sample selection effects, AGN classification ambiguities, and bar detection create difficulties for the study of the bar--AGN connection, on the other hand, and despite clear evidence of bars building up a gas reservoir at galaxy centres, theoretical caveats may make the detection of a correlation between both phenomena difficult.

Despite the presence of a bar, the galaxy needs to have gas available, which creates a secondary parameter. In addition, the roles of inner spiral arms and rings, inner bars, and dynamical resonances near the centre are not easy to disentangle in this process. Finally, AGN are stochastic phenomena, and the typical timescale for funnelling the gas to the centre of the galaxy (10$^8$ yr) and the lifetime of bars ($\sim$ 10$^{10}$ yr) are expected to be much longer than the timescale for AGN activity \citep[10$^7$ yr,][]{combes01}.

Given all these difficulties and different approaches in the effort to reduce bias, it is not surprising that different studies find conflicting results. 
Building on previous studies, and in an attempt to address the most important selection biases in a robust way, in this work we investigate the bar--AGN connection in a sample of galaxies from \citet[\citetalias{gadotti+09} hereafter]{gadotti+09} obtained from the SDSS. 
We use a new approach to match samples of active and inactive galaxies so that some of their properties have similar distributions in both groups.

A common approach in studying a bar--AGN connection is to use binomial population proportions, that is, to analyse the fraction of the population of galaxies that host AGN in samples of barred and unbarred galaxies. In addition to this approach, we use properties that are not dichotomous, such as bar strength and AGN intensity \citep[e.g.][]{cisternas+13}. This allows us to directly compare properties of the bar with those of the AGN, in an attempt to find a link between these two phenomena. 
In \S~2 we describe the sample of galaxies used in this work and in \S~3 the method used to define the control sample. In \S~4 we present the results of the search for a connection between bars and AGN and in \S~5 and \S~6 we present a discussion and our conclusions, respectively. Throughout this work we adopt a cosmology with $H_0=75$ km s$^{-1}$ Mpc$^{-1}$, $\Omega_{M} =0.3$, and $\Omega_{\Lambda}=0.7$.

\section{Sample description}

In this work we used subsets of the galaxies studied by \citetalias{gadotti+09}, obtained from all objects identified as galaxies in the SDSS Data Release 2 \citep{abazajian+04}. The bar and galaxy properties were measured using \texttt{BUDDA}\footnote{\url{www.sc.eso.org/~dgadotti/buddaonsdss.html}} \citep{desouza2004}, a code that performs 2D bulge--bar--disc decompositions using g, r, and i-band images.

From the SDSS data, \citetalias{gadotti+09} used only objects with stellar masses greater than $10^{10}$ M$_{\odot}$, excluding dwarf galaxies from the analysis. The redshift was restricted to 0.02~$\leq$~z~$\leq$~0.07 in order to guarantee a physical spatial resolution that would allow the study of the properties in an appropriate way. All objects present in the sample are seen at a face-on projection, with a semi-minor to semi-major axial ratio greater than or equal to 0.9. Galaxies in mergers or with perturbations caused by another nearby galaxy were excluded; objects that have some distortion in the CCD or a nearby bright star and galaxies with semi-major axis $a<4~\textrm{arcsec}$ were also excluded. These criteria provided a sample that is both representative and appropriate for two-dimensional bulge--disc--bar decomposition, because selecting face-on galaxies minimises dust and projection effects and facilitates the identification of bars. Given these constraints, the final sample provided by \citetalias{gadotti+09} is composed of 946 galaxies. The redshift of most objects in that sample is around 0.04~$\leq$~z~$\leq$~0.06, resulting in a typical spatial resolution of about 1.5~kpc.

Employing \texttt{BUDDA}, in \citetalias{gadotti+09} the images were decomposed using a two-dimensional model composed of an exponential disc, a Sérsic bulge, and a bar characterised by concentric generalised ellipses, as described in \cite{Athanassoula1990}:
 \begin{equation}
     \left( \dfrac{\left|x\right|}{a} \right)^{c} + \left( \dfrac{\left|y \right|}{b} \right)^{c} = 1 \, ,
     \label{eq:fit_bar}
 \end{equation}
where $a$ and $b$ correspond respectively to the semi-major and semi-minor axes, $x$ and $y$ to the coordinates of the pixels, and $c$ to the parameter that forms the ellipse. For $c > 2,$ the ellipse has a boxy shape which is more suitable for the description of the bar isophotes. Modeling the galaxy with multiple components, as done with \texttt{BUDDA}, provides the properties of the bar with more accuracy than when modeling the image using only ellipse fitting, with no decomposition. Fitting the image exclusively with ellipses tends to underestimate bar ellipticity \citep[see][]{gadotti2008}. For all galaxies in the sample, the detection of the bar and the measurement of its properties was done using this uniform method. 

As mentioned in \citetalias{gadotti+09}, most bars smaller than 2 - 3 kpc were probably missed. To circumvent this limitation and its possible associated bias, we adopted the same procedure as \cite{coelho+11}, excluding from the sample the galaxies with a bulge-to-total-luminosity ratio (B/T) of less than 0.043. As found in \cite{graham2008}, this ratio is related to galaxies later than Sc where smaller bars are more common.

The data in our galaxy sample were complemented by those provided for active objects of the SDSS DR2 studied by the MPA/JHU team\footnote{\url{wwwmpa.mpa-garching.mpg.de/SDSS/index_dr2.html}}. These data were described in \cite{Kauffmann2003} or obtained directly from the Catalog Archive Server Jobs System 
(CasJobs)\footnote{\url{skyserver.sdss.org/casjobs/}}.

\begin{figure}
    \centering
    \includegraphics[width=\linewidth]{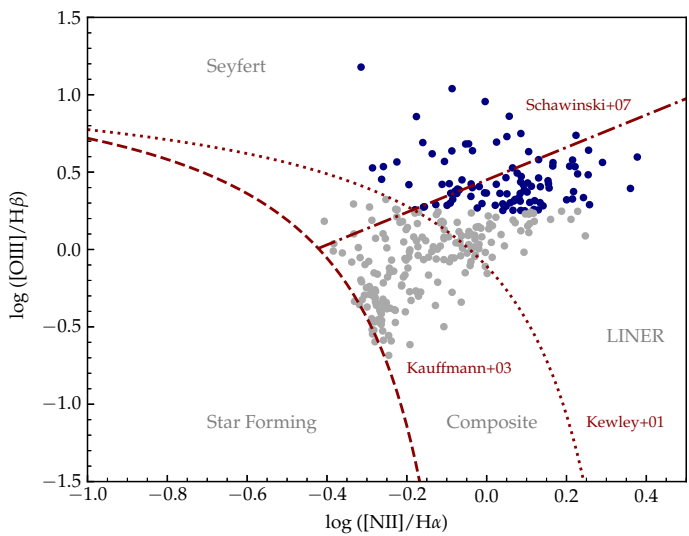}
    \caption{BPT diagram \citep{baldwin1981} for classification of objects with respect to their activity. The classification is obtained given the location of the objects in the diagram. The regions are marked according to the curves defined by \citet{Kewley2001}, \citet{Kauffmann2003}, and \citet{Schawinski2007} which are indicated in the diagram. The line ratios are described in \citet{Kauffmann2003}.}
    \label{fig:bpt}
\end{figure}

The galaxies in our sample were divided into active and inactive galaxies. For the sample of active galaxies, we selected all galaxies classified as AGN by \citet {Kauffmann2003}. We then used the BPT diagram \citep{baldwin1981} ---based on the $[\ion{O}{iii}]5007$/H$\beta$ and $[\ion{N}{ii}]6584$/H$\alpha$ ratios provided by \cite{Kauffmann2003}--- to refine this subsample. Fig.~\ref{fig:bpt} shows a BPT diagram containing the objects of our active subsample, where some distinct regions can be seen: below the curve proposed by \cite{Kauffmann2003} are the objects classified as starbursts. Between this curve and the one proposed by \cite{Kewley2001} are the objects now considered to be compounds or `transition region objects'. On the right side of the diagram and below the curve proposed by \cite{Schawinski2007} are the objects classified as low-ionization nuclear emission-line region (LINER). The last region, located at the top of the BPT diagram, is the region containing Seyferts.

We used the BPT diagram to remove composites from the AGN sample. 
To remove LINERs from the sample that might have their origin from mechanisms other than an AGN \citep{Heckman1987} the $[\ion{O}{iii}]5007$/H$\beta$ was used as a proxy for activity. Only objects with $\log \left([\ion{O}{iii}]5007/\text{H}\beta \right) > 0.25$ were kept in the sample of active objects.
The final AGN sample is shown in blue in Fig.~\ref{fig:bpt}. After this additional cut based on the $[\ion{O}{iii}]5007/\text{H}\beta$, and the adoption of the lower limit of B/T, the number of galaxies in the sample becomes 524, of which 94 are classified as active. This sample of galaxies represents the main sample of this work. Table~\ref{tab:sample_size} summarises the sample cuts and sizes, including the samples after the match procedure described in Sect.~\ref{sect:matching}.

\begin{table}
\centering
\caption{Sample size after applying constraints.}
\label{tab:sample_size}
\begin{tabular}{llc}
\hline
Sample                                                                                & Constraints                                                  & Total                \\ \hline
                                                                                      & b/a $>$ 0.9                                                  &                      \\
G09                                                                                   & 0.02 $\leq$ z $\leq$ 0.07                                    & 946                  \\
                                                                                      & a $>$ 4 arcsec                                               &                      \\  \hline
This work main sample,                                                                & B/T (lum.) $> 0.043$                                         & \multirow{2}{*}{524} \\
without matching                                                                      & $\log \left([\ion{O}{iii}]/\text{H}\beta \right) > 0.25^{a}$     &                      \\  \hline
Matched by M$_{\star}$,  M$_{\star, \rm{bulge}}$                                      & the same constraints                                         & \multirow{2}{*}{\centered{188}} \\               
and $\Sigma_{5}$                                                             & of the main sample                                           & \\  \hline
Matched by M$_{\star}$, M$_{\star,\rm{bulge}}$,                                       & the same constraints                                         & \multirow{3}{*}{\centered{188}} \\
Sérsic index of bulge,                                                                & of the main sample                                           &                                  \\
and $\Sigma_{5}$                                                             &                                                              &                                  \\  \hline
Matched by M$_{\star}$, M$_{\star, \rm{bulge}}$,                                      & the same constraints                                         & \multirow{2}{*}{188} \\
$\Sigma_{5}$ and g-r                                                         & of the main sample                                           &        \\  \hline             
\end{tabular}
\tablefoot{We note that although the number of galaxies is the same in the matched samples, they do not contain exactly the same galaxies. $^a$~Criterion adopted for active galaxies in the sample.}
\end{table}

\subsection{Emission line detection at different redshifts}

Redshift may prevent the detection of the $[\ion{O}{III}]$ emission lines with SDSS spectroscopy in more distant galaxies. This is likely to mainly affect galaxies hosting faint AGN where the emission lines could be undetected, leading the galaxy to be misclassified as inactive. When comparing for example the fraction of AGN in barred and unbarred galaxies, the possible inclusion of galaxies with an existing but undetected AGN (thus considered inactive) should affect both groups of barred and unbarred galaxies. For higher redshift, the fraction of the galaxy encompassed by the fibre is larger and the effect can be significant. When the $[\ion{O}{iii}]$ luminosity is directly compared to galaxy and bar properties, we apply a normalisation of the $[\ion{O}{iii}]$ luminosity by the fibre luminosity to minimise these possible effects as argued in \citet{oh+12}.

\begin{figure}
    \centering
    \includegraphics[width=\linewidth]{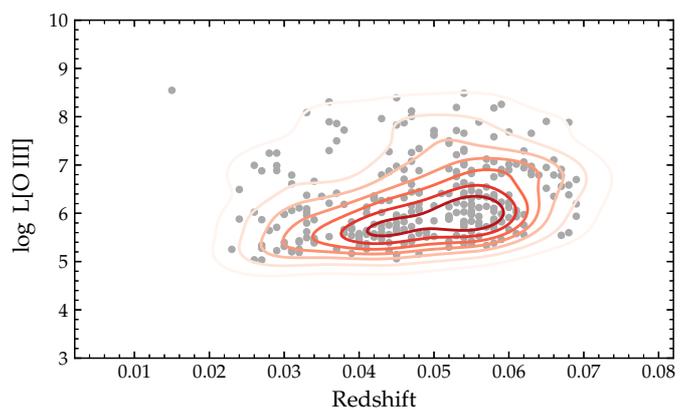}
    \caption{Logarithm of $[\ion{O}{iii}]$ luminosity provided by \citet{Kauffmann2003} of galaxies in our sample hosting an AGN as a function of redshift. Most galaxies are constrained in a range of redshift of 0.02~$\leq$z$\leq$~0.07.}
    \label{fig:oiii_vs_z}
\end{figure}

In Fig.~\ref{fig:oiii_vs_z}, it is possible to observe that there is no considerable decline in the detectable level of $[\ion{O}{III}]$ as a function of the small redshift interval in which the galaxies contained in our sample are located. For z $\leq$ 0.03, the lowest value of detected $\log$[\ion{O}{iii}] is 5.03 while for z $\geq$ 0.06 it is 5.54. Thus, the analysis presented in this work should not be strongly affected by issues related to redshift.

Dependence on the area encompassed by the fibre can also affect velocity dispersion measurements. In this case, we apply a correction for the velocity dispersion according to Eq.~(1) in \citet{cappellari2006} as done in \citet{oh+12}:
\begin{equation}
    \left(\dfrac{\sigma_{R}}{\sigma_{e}}\right) = \left(\dfrac{R}{R_{e}} \right)^{-0.066 \pm 0.035} \, , \label{eq:corr_vdisp}
\end{equation}
where $R$ is the radius comprised by the SDSS $1.5^{\prime\prime}$ fibre, $\sigma_{R}$ and $\sigma_{e}$ are the measured velocity dispersion and corrected velocity dispersion at one effective radius of galaxy ($R_{e}$), respectively. The velocity dispersion is used in determining the SMBH mass, applying the M--$\sigma$ relation below.

\subsection{Supermassive black hole mass estimation}
\label{subsec:blackhole}

To estimate the SMBH mass, we employ the M--$\sigma$ relation which was found independently by \cite{ferrarese2000} and \cite{gebhardt2000}:
\begin{equation}
    \log \left( \dfrac{M_{\bullet}}{M_{\odot}} \right) = \alpha + \beta \log \left( \dfrac{\sigma_{e}}{200 \, \text{km s}^{-1}} \right) \, ,
\end{equation}
where $M_{\bullet}/M_{\odot}$ is the SMBH mass in solar units, $\sigma_{e}$ is the corrected velocity dispersion at one effective radius in km s$^{-1}$ given by Eq.~\ref{eq:corr_vdisp}, and $\alpha$ and $\beta$ are the resulting regression coefficients in samples where the black hole mass is estimated by observations of the SMBH sphere of influence \citep[see][for more detailed discussion and others methodology description]{ferrarese2000, gebhardt2000}.

Later works, such as that by \cite{graham+11}, claim a reduction in scattering of M--$\sigma$ when a sample of galaxies is divided according to morphology. In particular, when a sample is divided into subsamples of barred and unbarred galaxies, such a reduction in scattering is also noticed. These latter authors suggested a hypothesis to explain such an effect whereby bars enhance the central velocity dispersion of a galaxy as they evolve. Hereafter, unless indicated, we use the following relations to estimate the SMBH mass \citep{graham+11}. For unbarred galaxies we apply
\begin{equation}
    \log \left( \dfrac{M_{\bullet}}{M_{\odot}} \right) = (8.25 \pm 0.06) + (4.57 \pm 0.35) \log \left( \dfrac{\sigma_{e}}{200 \, \text{km s}^{-1}} \right) \, ,
    \label{eq:m-sigma_nao-barrada}
\end{equation}
while, for barred galaxies we will employ
\begin{equation}
    \log \left( \dfrac{M_{\bullet}}{M_{\odot}} \right) = (7.80 \pm 0.10) + (4.34 \pm 0.56) \log \left( \dfrac{\sigma_{e}}{200 \, \text{km  s}^{-1}} \right) \, .
    \label{eq:m-sigma_barrada}
\end{equation}
When using a common relation for barred and unbarred galaxies, for comparison purposes, we use the relation obtained for the complete sample of \cite{graham+11}, without distinction of morphological aspects
\begin{equation}
    \log \left( \dfrac{M_{\bullet}}{M_{\odot}} \right) = (8.13 \pm 0.005) + (5.13 \pm 0.34) \log \left( \dfrac{\sigma_{e}}{200 \, \text{km  s}^{-1}} \right) \, .
    \label{eq:m-sigma_common}
\end{equation}

\subsection{Environmental density}

In a previous study, \cite{alonso14}, using an AGN sample, found that the fraction of barred AGN in groups and clusters is higher than the fraction of barred AGN overall. This result suggests an important role of environment in the process of AGN feeding, which we must account for in a scenario where the bars are part of the mechanism that feeds the black holes. As a proxy for the environmental density of the galaxies in the sample, we estimated the projected local density using
\begin{equation}
    \Sigma_{n} = \dfrac{n}{\pi d_{n}^{2}} \, , 
    \label{eq:env_density}
\end{equation}
where $d_{n}$ is the projected distance to the n-th nearest neighbouring galaxy. In computing $\Sigma$, we consider only neighbouring galaxies with $M_{r}<-20.5$ as applied by \cite{alonso13} and with a radial velocity interval given by $|\Delta z c|~<~1000~\text{km s}^{-1}$ \citep{ baldry06, alonso13}.

In this way, we estimate $\Sigma_{5}$ which has a median of $\Sigma_{5}~=~0.32~\text{Mpc}^{-2}$ for the total sample of 524 galaxies as well as quartiles Q1 and Q3 respectively equal to $0.21~\text{Mpc}^{-2}$ and $0.51~\text{Mpc}^{-2}$. In addition, we did not find significant differences between the $\Sigma_{5}$, $\Sigma_{4}$, and $\Sigma_{4,5}$ distributions, where $\Sigma_{4,5}$ is the average for n = \{4, 5\}. The comparison between these distributions with the k-sample Anderson-Darling test \citep{Scholz1987} presents a p-value > 0.25. Considering the values reported by \citet{baldry06}, where $\Sigma$ ranges from 0.05~\text{Mpc}$^{-2}$ to 20~\text{Mpc}$^{-2}$ for galaxies from void environments to groups, respectively, the galaxies in our sample can be considered to be in environments of relative isolation. $\Sigma_{5}$ will be employed in the subsequent analysis in order to compare active and inactive galaxy samples with similar $\Sigma_{5}$ distributions.

\subsection{Bar properties and morphological classification}

The properties of the bars are related to the placement of the galaxy in the Hubble sequence. In general, bars  are larger in earlier spirals than in later spirals \citep[relative to the galaxy's 25~mag~arcsec{$^{-2}$} isophote;][and references therein]{sellwilkin1993}. In addition, some studies have indicated that the fraction of bars may also be dependent on the morphological classification \citep{lee2019}, suggesting a bimodal distribution in the fraction of bars between earlier and later type spirals.

The T-type may be used to divide spiral galaxies between earlier and later types. In recent years, considerable progress has been made in expanding the morphological classification catalogues available for SDSS galaxies through machine learning techniques, providing the T-type for an increasing number of objects \citep[e.g.][]{dominguez+18}. Still, these catalogues should only cover part of the galaxies in our sample.

Given the restriction on the ratio between bulge luminosity and total luminosity that we applied to our sample, following the relation found by \citet{graham2008}, most of the galaxies later than Sc in the Hubble sequence are excluded from our sample. However, the classification made by this parameter has a spread. This means that, despite having mostly earlier spiral galaxies, some later spirals may still be present in the sample. Even so, the additional matching processes carried out using bulge properties (M$_{\star}$ and Sérsic index of the bulge) and galaxy colour (g-r corrected for extinction) should provide an additional reduction in this possible associated bias, because these properties also correlate with morphology. As in all the possible biases we explore above, this whole set of efforts can contribute to reducing potential biases. In the following section, we describe the matching methodology as an extra effort to minimise bias.

\section{Description of the matching technique}
\label{sect:matching}

In this work, we want to compare the properties of bars in active and inactive galaxies. To this end, distributions of the global properties of the galaxies in these two subsamples should be as similar as possible in order to minimise selection bias.
A matching process can then be used to achieve this goal, randomising distributions of parameters of a set of selected confounding covariates. In this work, we use the \texttt{R} package \texttt{MatchIt} \citep{ho2007matching} to implement the matching process. \texttt{MatchIt} is based on propensity score matching (PSM), introduced by \cite{rosenbaumpsm}. 
PSM has been employed before in astronomy. For example,  \citet{rafaelpsm} used PSM to build a control sample of inactive galaxies in order to study the effect of cluster environment on galaxy activity, and \citet{dantas+21} used PSM to compare the global properties of red sequence galaxies with and without the UV upturn phenomenon.

The propensity score $e(x)$ provides the conditional probability of an object to be designated as part of the test or control groups given a set $x$ of selected covariates. In the case of this study, the probability of a galaxy being active ($Z = 1)$ or inactive ($Z=0$) given the covariates is:
 \begin{equation}
     e(x) = P(Z = 1 | x)\, .
     \label{eq:psm}
 \end{equation}
Only in very particular circumstances do we know the exact propensity score; in all other cases, it needs to be estimated. In the case of a dichotomous variable, for example active versus inactive, given a set of covariates, PSM can be estimated from a model where the PSM logit is a linear function of the cofactors where $\boldsymbol {\beta} = \lbrace \beta_{1}, \beta_{2}, ..., \beta_{k} \rbrace$ is the vector of the regression coefficients:
\begin{equation}
     \textrm{logit}({e_i}) = \ln \frac{e_i(x)}{1-e_i(x)} =  \boldsymbol{\beta}\boldsymbol{x}^T \,,
\end{equation}
and then
\begin{equation}
     e_{i}(x) = \frac{1}{1 + e^{-\boldsymbol{\beta}\boldsymbol{x}^T}} \,,
     \label{eq:psm_estm}
\end{equation}
where $\boldsymbol{x}^T$ is the transpose of the cofactor vector.

The PSM becomes the property with which the pairs of active and inactive objects will be created after estimating the PSM for each sample object (see \cite{stuart2010} for a review of matching methods). The simplest approach is to measure the distance based on the absolute difference between the PSM $(D_{ij} = |e_{i} - e_{j}|) $.

\texttt{MatchIt} offers several options to search for objects in the test group for building pairs \citep[see][for all options]{ho2007matching}. In this case, as the group of active objects is smaller than that of inactive objects, the k-nearest neighbour algorithm (kNN) achieves a good performance in the matching process. Even though the optimal matching methods seek to minimise the distance between the pairs globally, in the scenario where there is little competition for pairs, the kNN and optimal matching should not present great divergences. As presented in \cite{gu1993}, the control units selected by kNN and optimal matching tend to be the same in general, which does not significantly change the distributions. The only difference is in the way the pairs are associated. In optimal matching, the difference between the propensity 
score in pairs is minimised but the overall effect turns out to be subtle when there is little competition for control units. In our sample, there are more than two objects in the control sample for each object in the treatment sample.

After the match, it is necessary to ensure that the selected variables are balanced between the control and test groups. As seen in \cite{stuart2010}, graphical or numerical diagnostics can be used. In an ideal scenario, $\widetilde{p} (x \vert Z=1) = \widetilde{p} (x \vert Z=0)$, where $\widetilde{p}$ express the empirical distribution. However, this ideal outcome is difficult to achieve. Between numerical diagnoses, standardised difference in means (SDM) is one of the most common \citep{stuart2010, king2011matchit}:
\begin{equation}
    \textrm{SDM} = \dfrac{\overline{x}_{k,t}-\overline{x}_{k,c}}{\sigma_{k,t}} \,  ,
    \label{eq:sdm}
\end{equation}
where $\overline{x}_{k,c}$ and $\overline{x}_{k,t}$ are, respectively, the mean of the distribution of covariate $k$ for the treatment and control groups and $\sigma_{k,t}$ is the standard deviation of covariate $k$ in the treatment group. \texttt{MatchIt} includes routines that perform these computations after an additional matching process.

According to \cite{rubin2001}, SDM must be less than 0.25 for the balance of covariates between the control and test groups to be trustworthy. As shown in Table~\ref{tab:sdm}, all our matches satisfy this requirement. For graphical diagnostics there are multiple possibilities, for example, histograms or distributions of the propensity score, a QQ-plot of the covariate distributions, or even an SDM plot before and after a match \citep[][and references therein for examples]{stuart2010}. 

After the selection steps described in the previous section, we are left with a sample of galaxies composed of 209 barred and 315 unbarred galaxies. We then performed three different matches, based on combinations of total and bulge stellar mass, $\Sigma_{5}$, Sérsic index of the bulge, and colour g-r (the latter is computed with the modelled magnitudes corrected for extinction provided by SDSS). The matches are done as follows: for each object classified as active, an object is chosen from the pool of inactive galaxies that is as similar as possible to the active galaxy in terms of its properties based on the propensity score of the matching. In this way, there is always a one-to-one correspondence between the active and control samples.
The selection of these variables to be randomised was based on their potential to represent biases to measure the effect of the bar on the feeding process. 

\begin{figure}
    \centering
    \includegraphics[width =\linewidth]{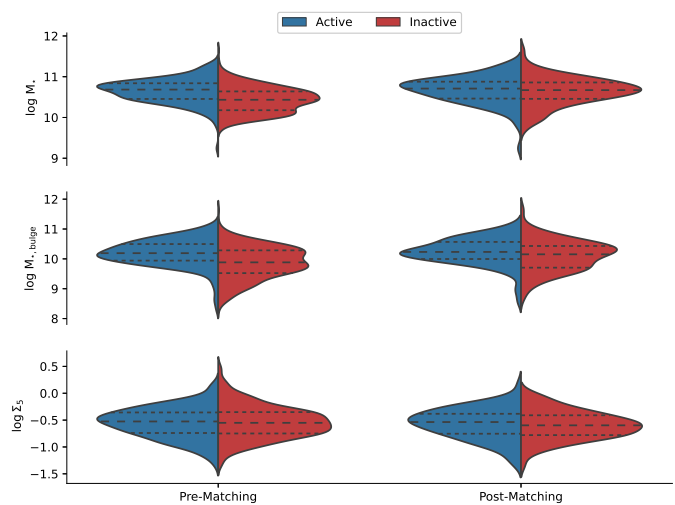}
    \caption{Matching graphical diagnostic similar to that found in \citet{rafaelpsm} for M${_\star}$, M$_{\star,\rm{bulge}}$, and $\Sigma_{5}$ (simultaneous matching). The figure shows the distributions of active and inactive objects before and after the matching process. Long-dashed and short-dashed lines show the mean value of the distribution and the quartiles Q1 and Q3, respectively.}
    \label{fig:match_mt_mb}
\end{figure}

AGN are preferentially found in massive galaxies \citep{Kauffmann2003}, and so the first general galaxy property we must randomise is the total stellar mass. Also, there is a correlation between the bulge mass (or luminosity) and the black hole mass (M$_{\bullet}$) \citep{ferrarese2000, gebhardt2000}. Matching the samples by total and bulge mass (M$_\star$ and M$_{\star,\rm{bulge}}$) guarantees we are comparing galaxies with the same mass distribution and nuclear properties. Matching samples using the bulge mass is often overlooked given the uncommon availability of structural decompositions. M$_{\star}$ is the sum of the masses of each of the components, which in turn were determined individually. With \texttt{BUDDA}, the fits are performed in several bands, allowing colours to be estimated in each component in \citetalias{gadotti+09}. These colours serve as a basis for estimating the mass-to-light ratio ($\mathcal{M}/L$). The mass estimate of each component is then the result of the mass-to-light ratio estimate. In addition to the bulge and total mass, we also matched the sample by $\Sigma_5$ in order to take into account the environment. To visualise the result of the matching process, we present a graphical diagnostic similar to that found in \cite{rafaelpsm}, with the distributions of covariates before and after matching (Fig.~\ref{fig:match_mt_mb}), where a noticeable improvement in the proximity of the averages and quartiles (Q1 and Q3) can be seen. The SDM obtained according to Eq.~\ref{eq:sdm} in this matching and in the subsequent ones are shown in Table~\ref{tab:sdm}.

\begin{figure}
    \centering
    \includegraphics[width =\linewidth]{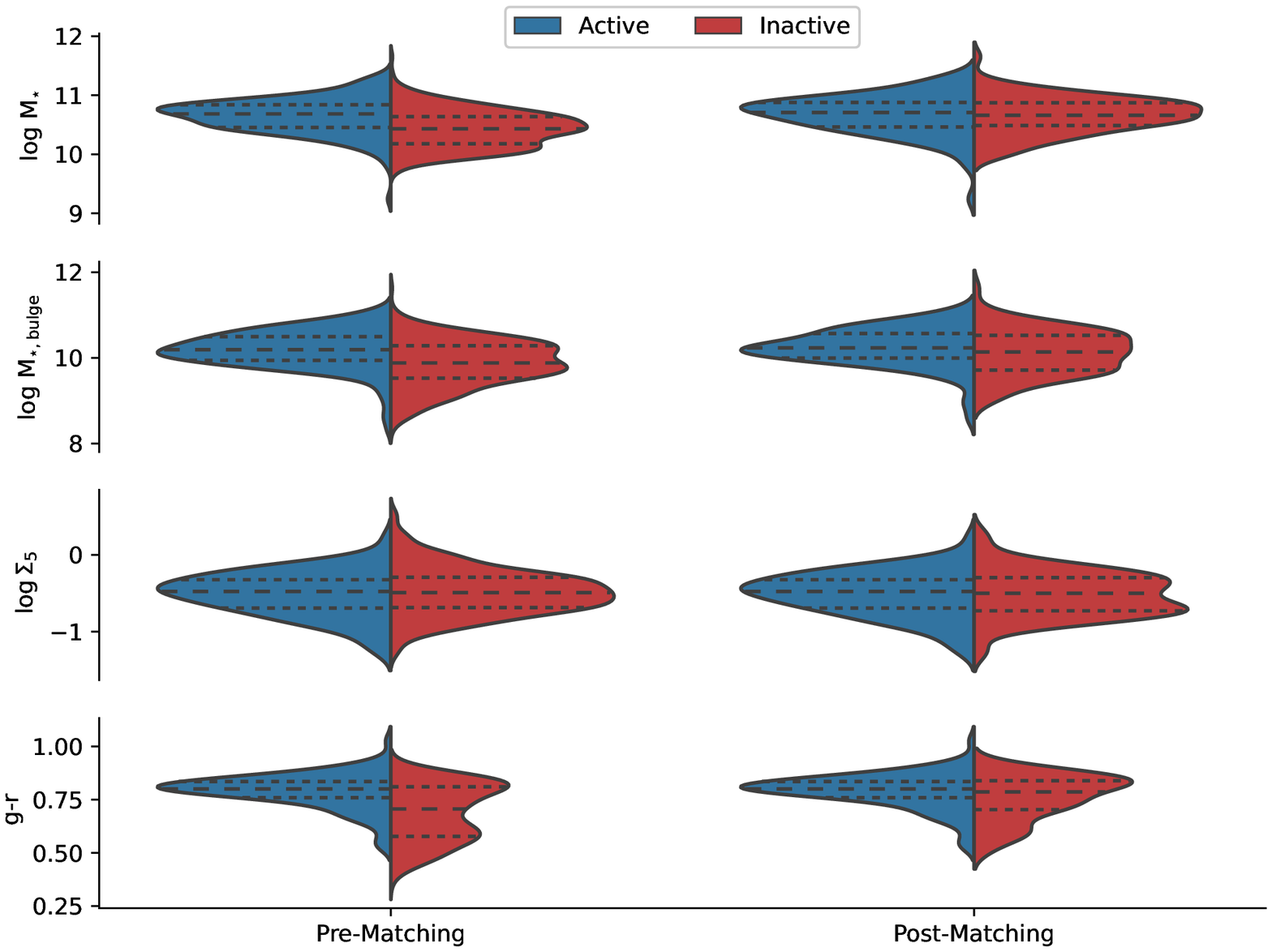}
    \caption{Matching graphical diagnostic for M${_\star}$, M$_{\star, \rm{bulge}}$, $\Sigma_{5}$, and g-r (simultaneous matching). The figure shows the distributions of active and inactive objects before and after the matching process. Lines are the same as in Fig.~\ref{fig:match_mt_mb}.}
    \label{fig:match_mt_mb_cor}
\end{figure}

The optical colours of the galaxies are correlated with their morphological type. There may be variations
in the frequency of bars with colour, although this is still being debated. 
\cite{oh+12} find that most of the unbarred spirals are located in the blue cloud, whereas barred spiral galaxies are redder and brighter than typical late types, and concentrated in the green valley. Similar results are reported by \citet{masters2011} who identified bars in 70\% of their sample of red spirals in contrast with 25\% of the blue ones but in addition found that the fraction of bars correlates with bulge predominance as well. However, different results are found by \citet{erwin+18}, who finds that bars are as common in blue as in red galaxies. 
Another important factor to consider is the availability of gas for the AGN feeding. In a scenario where
bars are responsible for the momentum redistribution inside the galaxy, the presence of gas itself is an important factor to be considered. A study from \citet{lee2012} indicates that barred galaxies that have more activity tend to be bluer. 
Altogether, this shows that it is important to avoid colour biases in the study of bar effects.
Taking this into account, in addition to the matching of the stellar mass, the bulge mass, and the projected environmental density simultaneously, we perform a second round of matching that also considers the total galaxy colour (g-r). The distributions of covariates for this case are shown in Fig.~\ref{fig:match_mt_mb_cor}.

\begin{figure}
    \centering
    \includegraphics[width =\linewidth]{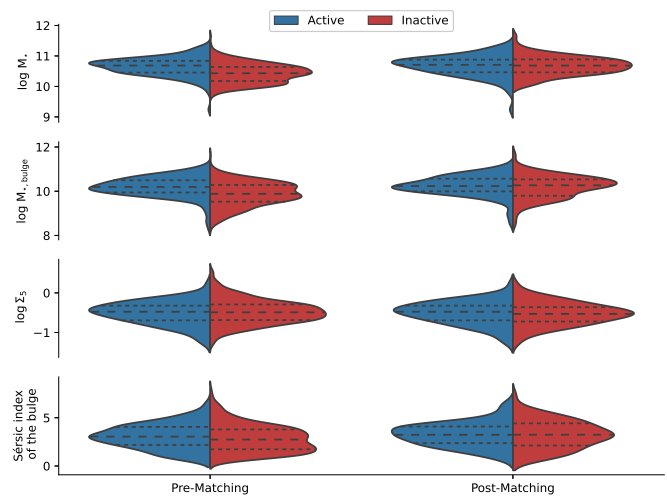}
    \caption{Matching graphical diagnostic for M${_\star}$, M$_{\star, \rm{bulge}}$, $\Sigma_{5}$, and Sérsic index of the bulge  (simultaneous matching). The figure shows the distributions of active and inactive objects before and after the matching process. Lines are the same as in Fig.~\ref{fig:match_mt_mb}.}
    \label{fig:match_mt_mb_sersic}
\end{figure}

Another important aspect to consider here is the presence and properties of a central bulge in the galaxies.
Today we know that the simple early view of galaxy bulges as scaled-down ellipticals is incomplete.
Disc galaxies might host classical bulges, which were presumably formed through violent processes with minor mergers producing hierarchical clustering, and what are referred to as pseudo-bulges formed over longer timescales via disc instabilities and secular evolution processes \citep[e.g.][]{wyse+97,kormendy+04, athanassoula05}. These different bulge categories are structurally distinct:
pseudo-bulges can either be peanuts, that is the vertically extended central part of the bar, or disc-like bulges. Disc-like bulges have properties similar to those of thickened discs, having younger stellar populations, kinematics supported by rotation, and a less concentrated surface brightness profile than 
classical bulges, which are characterised by a low Sérsic index
\citep[e.g.][]{carollo+97,gadotti+01,kormendy+06,fisher+08,gadotti+09}.
Disc-like bulges tend to be flatter than classical bulges, and indeed \citet{gadotti+20} showed that photometric bulges with exponential profiles or low Sérsic index and do not show a box or peanut morphology are in fact simply nuclear discs built by the bar.
As the process that builds nuclear discs can also build a reservoir of material that may feed AGN, the Sérsic index of the bulge can also be included in the matching process. We therefore performed a third round of the matching process, now considering M$_\star$, M$_{\star,\rm{bulge}}$, $\Sigma_{5}$, and the bulge Sérsic index. The distributions of covariates for this case are presented in Fig.~\ref{fig:match_mt_mb_sersic}.
These matched samples will be used in this work every time active and inactive galaxies are compared. 

\begin{table}
\centering
\caption{The SDM of the randomised properties in the matching process.}
\label{tab:sdm}
\begin{tabular}{lcc}
\hline\hline
\multicolumn{1}{c}{Property} & \begin{tabular}[c]{@{}c@{}} SDM before\\ matching\end{tabular} & \begin{tabular}[c]{@{}c@{}}SDM after\\ matching\end{tabular} \\ \hline
\multicolumn{3}{c}{Matched by M$_{\star}$, M$_{\star, \rm{bulge}}$, and $\Sigma_{5}$,}                                                                              \\ \hline
M$_{\star}$                  & 0.5984                                                    & 0.0466                                                   \\
M$_{\star, \rm{bulge}}$      & 0.4864                                                    & 0.0615
 \\
$\Sigma_{5}$                 &  -0.1092                                                  & 0.0946 \\ \hline
\multicolumn{3}{c}{Matched by M$_{\star}$, M$_{\star,\rm{bulge}}$, $\Sigma_{5}$, and Sérsic index of bulge}                                                        \\ \hline
M$_{\star}$                  & 0.5984                                                    & -0.0128                                                   \\
M$_{\star, \rm{bulge}}$      & 0.4864                                                    & -0.0984                                                  \\
$\Sigma_{5}$                 &  -0.1092                                                  & 0.0653
\\
Sérsic index of bulge        & 0.3425                                                    & 0.0177                                                \\ \hline
\multicolumn{3}{c}{Matched by M$_{\star}$, M$_{\star, \rm{bulge}}$, $\Sigma_{5}$, and g-r}                                                                         \\ \hline
M$_{\star}$                  & 0.5984                                                    & 0.0401                                                   \\
M$_{\star, \rm{bulge}}$      & 0.4864                                                    & 0.0004                                                  \\
$\Sigma_{5}$                 &  -0.1092                                                  & 0.0160
\\
g-r                          & 0.2476                                                    & -0.0409                                                 \\ \hline
\end{tabular}
\tablefoot{The SDM was obtained according to Eq.~\ref{eq:sdm} \citep{stuart2010, king2011matchit}. This measure serves as a means of evaluating the effectiveness of the matching process. A standardised difference of the means of less than 0.25 is an indicator of the effectiveness of this process \citep{rubin2001}. The SDM values are listed before and after the matching in each performance of this process for each of the properties.}
\end{table}

\section{Results}

In this section we present our search for signs of the bar--AGN connection. First we compare the activity in barred and unbarred galaxies. After that we show our search for correlations between bar properties and AGN strength.

\subsection{The fraction of AGN in barred and unbarred galaxies}

\begin{figure*}
\centering
\includegraphics[width=0.42\linewidth]{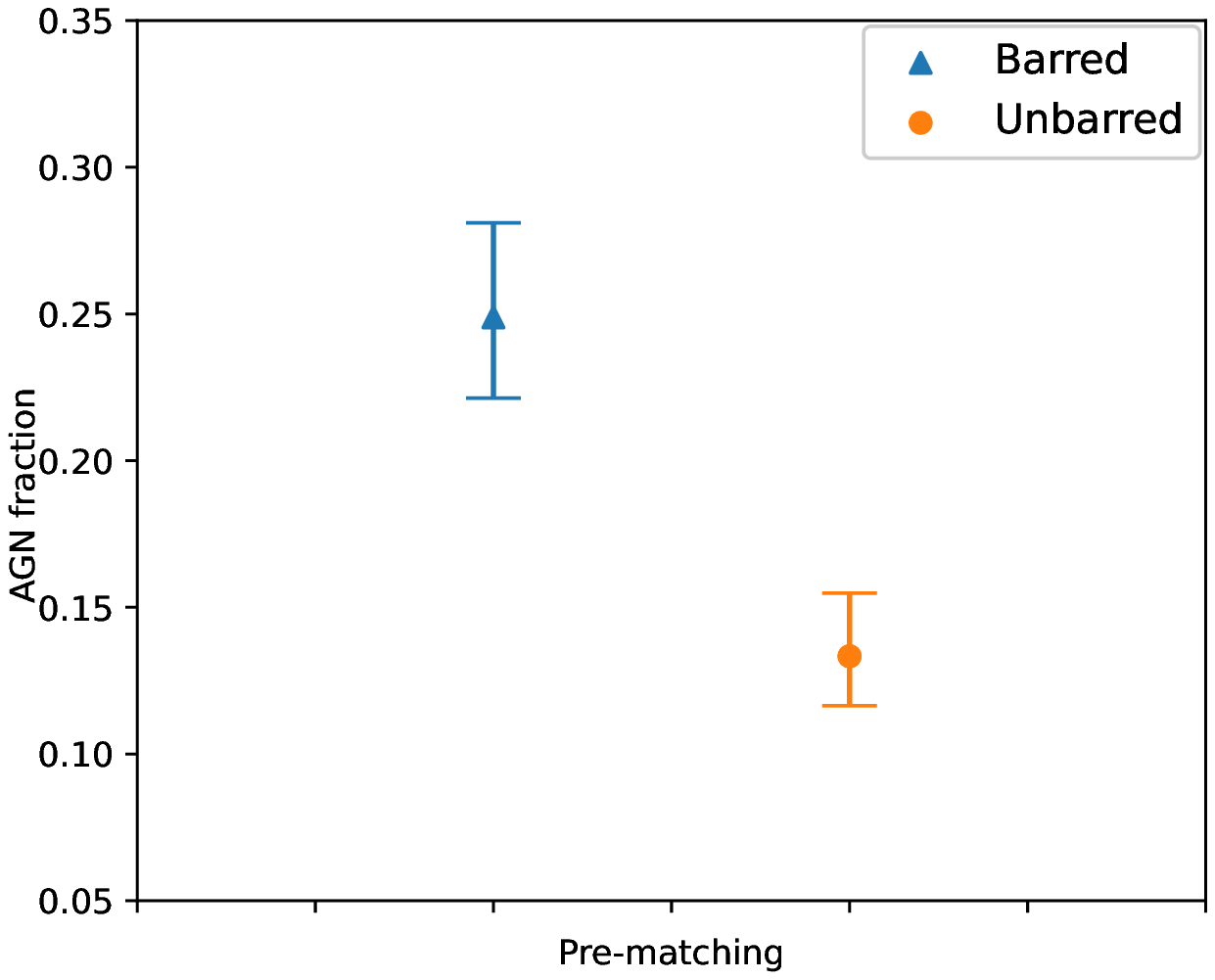}
\includegraphics[width=0.42\linewidth]{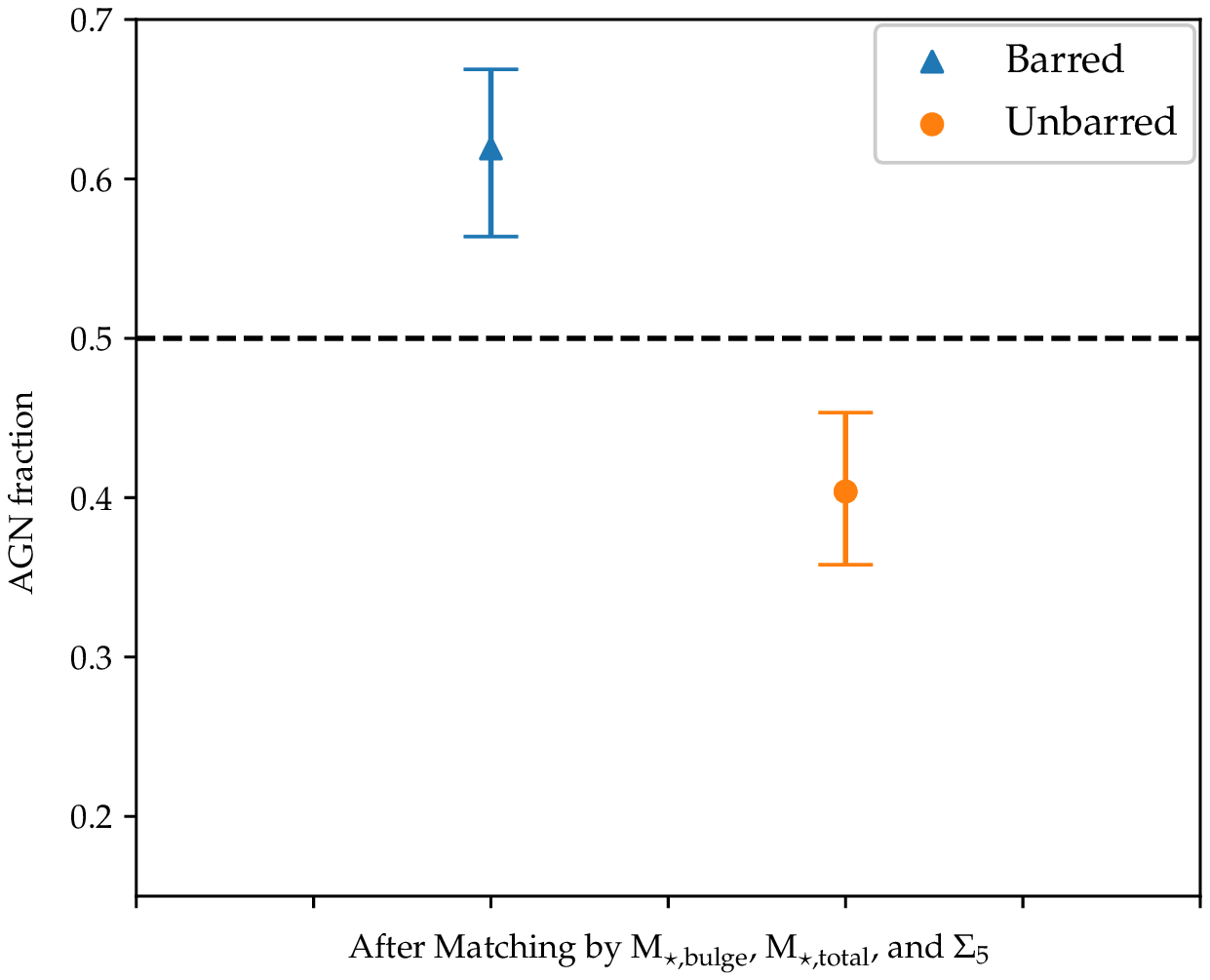}
\includegraphics[width=0.42\linewidth]{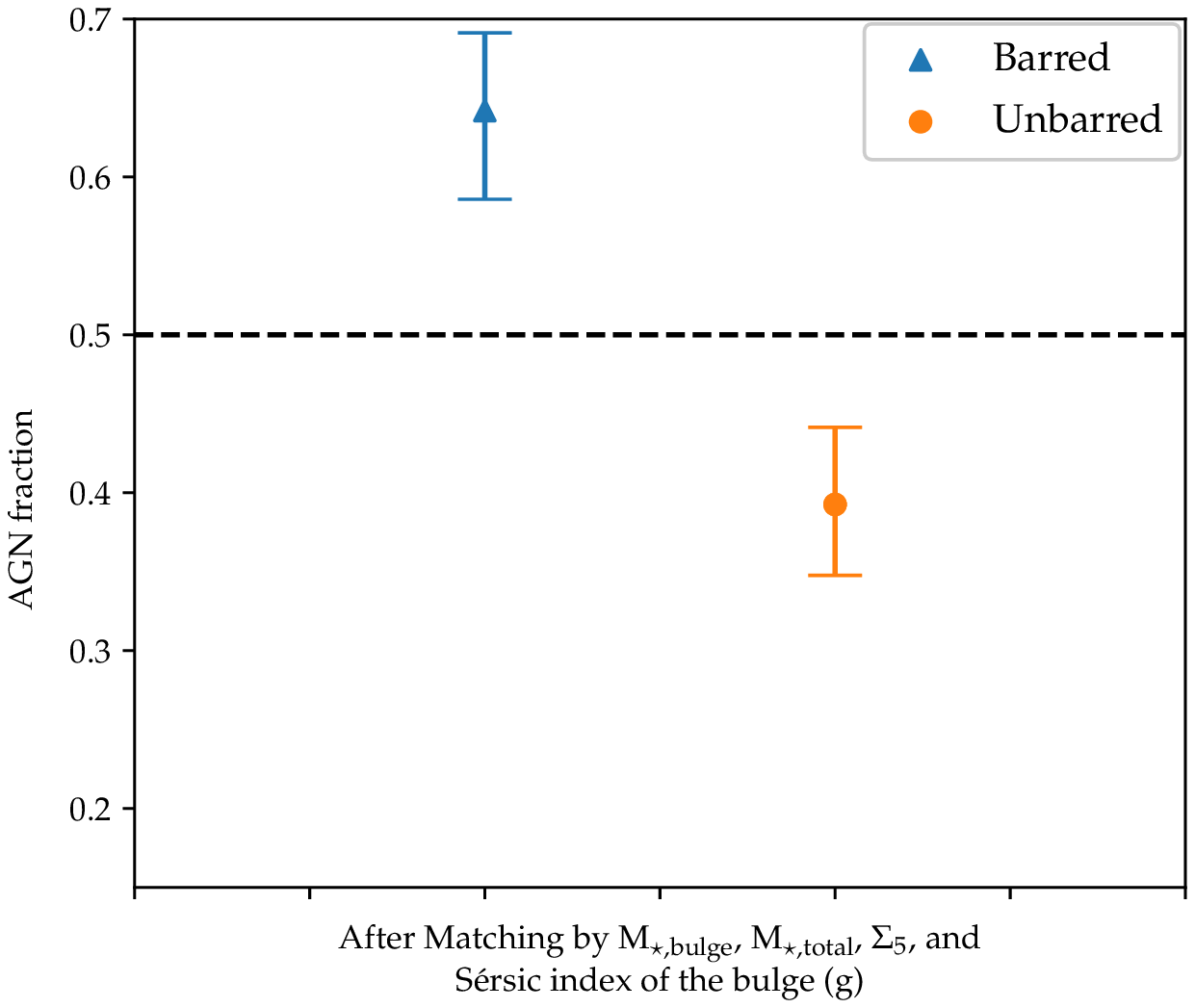}
\includegraphics[width=0.42\linewidth]{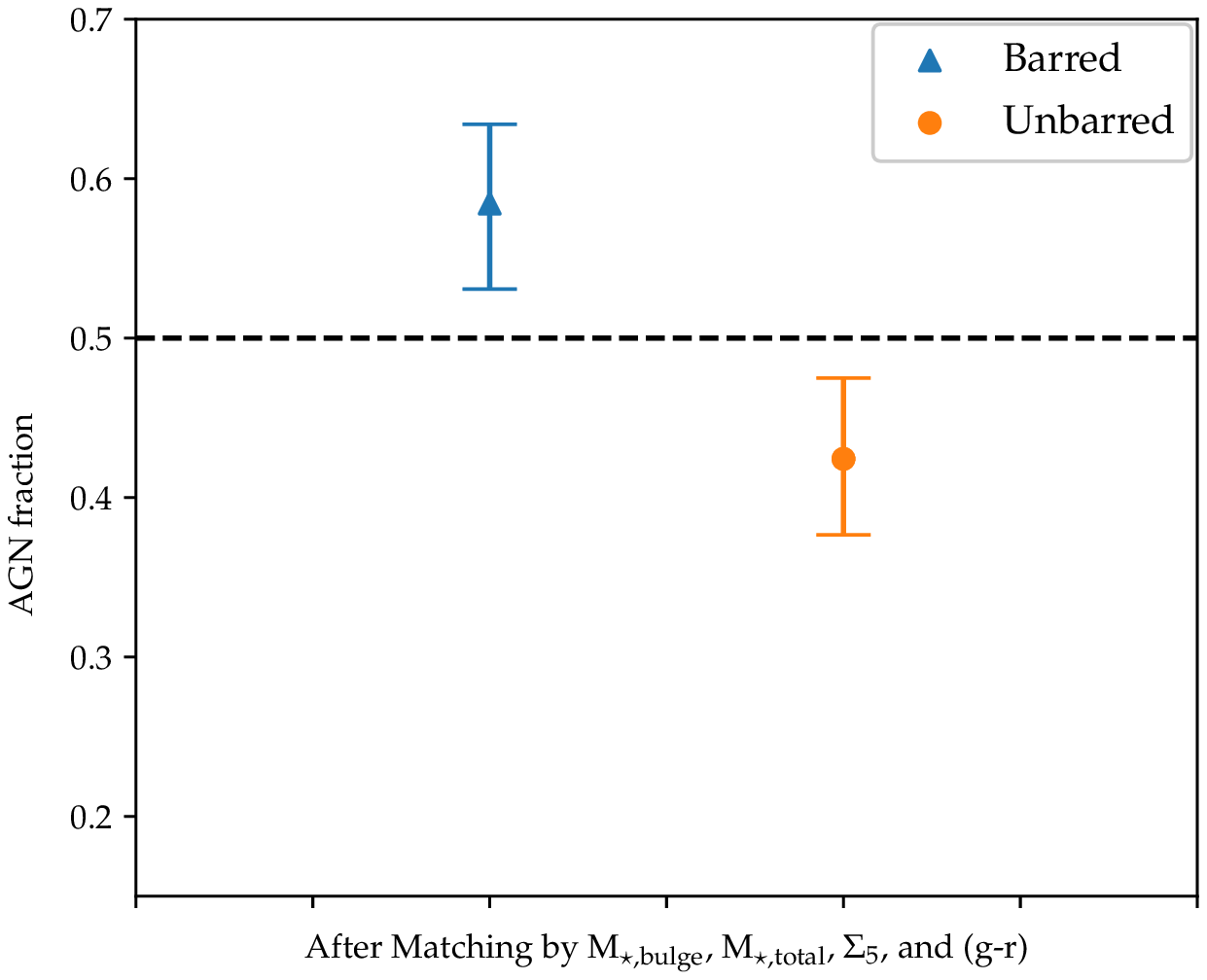}

\caption{Fraction of AGN in barred and unbarred galaxies for all objects of the sample. 
Top left: Sample before any matching. 
Top right: Sample after the matching process by M$_\star$, M$_{\star, \rm{bulge}}$ and $\Sigma_{5}$.
Bottom left: Sample after the matching process by M$_\star$, M$_{\star, \rm{bulge}}$, $\Sigma_{5}$ and colour (g-r).
Bottom right: Sample after the matching process by M$_\star$, M$_{\star, \rm{bulge}}$, $\Sigma_{5}$ and Sérsic index.
Error bars are 1$\sigma$ confidence level using the beta distribution \citep{cameron2011}.}
\label{figfrac}
\end{figure*}

The first test one can make is a straightforward comparison to check whether or not AGN are preferentially found in barred galaxies. First we use the sample matched by M$_{\star}$, M$_{\star, \rm{bulge}}$, and $\Sigma_{5}$. 
The AGN fraction in barred and unbarred galaxies for the sample defined after this matching process is shown in the top right panel of Fig.~\ref{figfrac}. In the left panel of this figure we show, for comparison, the test made with the whole sample (before the matching). 
Error bars were determined according to \citet{cameron2011}, who compares different methods applied in astronomy to identify confidence intervals. Implementation of this method proved to be quite simple because it uses the beta distribution. Furthermore, we obtained very similar results using the beta distribution in comparison with bootstrap \citep{efron79}. For the sake of simplicity, consistency with other methods, and also for the good performance in small samples \citep[see][]{cameron2011}, we employ the methodology proposed by Cameron in all subsequent confidence interval estimations. From Fig.~\ref{figfrac} it is possible to see that the global AGN fraction after matching, without distinction between barred and unbarred, is 50\%, which is indicated by the dashed lines in the figures. This is a result of the matching processes whereby for each active galaxy a inactive galaxy with similar properties is found. In both cases, AGN are preferentially found in barred galaxies. The higher fraction of AGN found in barred galaxies in comparison with the fraction of AGN found in unbarred galaxies was previously reported in the literature (e.g. \cite{galloway+15}. For the AGN fraction measured with the sample matched in M$_{\star}$, M$_{\star, \rm{bulge}}$ and $\Sigma_{5}$, the difference is of $2\sigma$.

As explained in Section 3, matching by galaxy colour is another important process which can be used to avoid bias due to morphology and gas availability. 
We now compare the fraction of AGN in barred and unbarred galaxies of the subsample created adding the match by colour (g-r). Results are shown in the bottom right panel of Fig.~\ref{figfrac}. Again, the fraction of AGN is larger in barred galaxies than in unbarred galaxies, now with a more significant difference than for the previous matched samples. Here the difference significance is $1.5\sigma$.

\begingroup
\setlength{\tabcolsep}{10pt} 
\renewcommand{\arraystretch}{1.5} 
\begin{table*}
\centering
\caption{Fractions of active galaxies.}
\label{tab:fraction}
\begin{tabular}{lccccccc}
\hline\hline
\multicolumn{1}{c}{Morphology} & All    & \multicolumn{2}{c}{Active} & \multicolumn{2}{c}{Seyfert} & \multicolumn{2}{c}{LINER} \\
\multicolumn{1}{c}{}           & number & number      & fraction     & number      & fraction      & number     & fraction     \\ \hline
\multicolumn{8}{c}{Without Matching}                                                                                           \\ \hline
Barred                         & 209    & 52          & $0.249\strut^{\,0.281}_{\,0.221}$            & 14               & $0.067\strut^{\,0.089}_{\,0.054}$             & 38              & $0.182\strut^{\,0.212}_{\,0.158}$        \\
Unbarred                       & 315    & 42          & $0.133\strut^{\,0.155}_{\,0.116}$            & 12               & $0.038\strut^{\,0.052}_{\,0.030}$             & 30              & $0.095\strut^{\,0.114}_{\,0.081}$        \\ \hline

\multicolumn{8}{c}{Matched by M$_{\star}$, M$_{\star, \rm{bulge}}$, and $\Sigma_{5}$}                                                              \\ \hline
Barred                         & 84     & 52          & $0.619\strut^{\,0.669}_{\,0.564}$      & 14          & $0.167\strut^{\,0.215}_{\,0.134}$           & 38         & $0.452\strut^{\,0.507}_{\,0.340}$        \\
Unbarred                       & 104     & 42          & $0.404\strut^{\,0.453}_{\,0.358}$      & 12          & $0.115\strut^{\,0.154}_{\,0.091}$           & 30         & $0.288\strut^{\,0.337}_{\,0.248}$        \\ \hline

\multicolumn{8}{c}{Matched by M$_{\star}$, M$_{\star, \rm{bulge}}$, $\Sigma_{5}$, and g-r}                                                         \\ \hline

Barred                         & 89     & 52          & $0.584\strut^{\,0.634}_{\,0.531}$      & 14          & $0.157\strut^{\,0.204}_{\,0.126}$           & 38         & $0.427\strut^{\,0.480}_{\,0.377}$        \\

Unbarred                       & 99    & 42          & $0.424\strut^{\,0.475}_{\,0.376}$      & 12          & $0.121\strut^{\,0.162}_{\,0.096}$           & 30         & $0.303\strut^{\,0.353}_{\,0.261}$        \\ \hline

\multicolumn{8}{c}{Matched by M$_{\star}$, M$_{\star,\rm{bulge}}$, $\Sigma_{5}$, and Sérsic index of bulge}                                        \\ \hline

Barred                         & 81     & 52          & $0.642 \strut^{\,0.691}_{\,0.586}$      & 14          & $0.173\strut^{\,0.223}_{\,0.139}$           & 38         & $0.469\strut^{\,0.525}_{\,0.415}$        \\
Unbarred                       & 107    & 42          & $0.392\strut^{\,0.441}_{\,0.348}$      & 12          & $0.112\strut^{\,0.150}_{\,0.088}$           & 30         & $0.280\strut^{\,0.328}_{\,0.241}$        \\ \hline
\end{tabular}
\tablefoot{Fraction of active galaxies among all the galaxies in the sample that have disk morphology and have a B/T > 0.043 in i-band of the SDSS \citep{graham2008}. Fractions are displayed for barred and unbarred galaxies. Fractions are recorded for samples without matching, for matching performed considering the sets of covariates \{M$_{\star}$; M$_{\star, \rm{bulge}}$; $\Sigma_{5}$\}, \{M$_{\star}$; M$_{\star, \rm{bulge}}$; $\Sigma_{5}$; g-r\} and  \{M$_{\star}$; M$_{\star, \rm{bulge}}$; $\Sigma_{5}$; Sérsic index of bulge\}.}
\end{table*}
\endgroup

Finally, we compared the subsample of galaxies matched by the Sérsic index of the bulge. The result for bar fraction in active and inactive galaxies in this case is shown in the bottom left panel of Fig.~\ref{figfrac}.
After matching by the Sérsic index of the bulge, the fraction of AGN is still larger in barred galaxies than in unbarred galaxies with significance of $2.3\sigma$. All these results are summarised in Table~\ref{tab:fraction}.

\begin{figure}
    \centering
    \includegraphics[width=\linewidth]{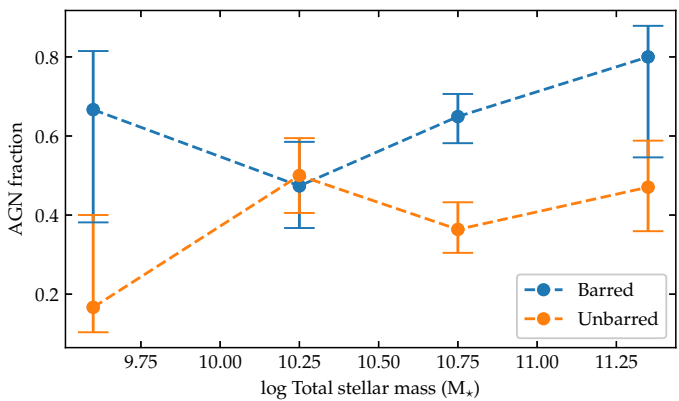}
    \caption{AGN fraction in barred and unbarred galaxies per mass bin. Here we use the sample matched by M$_\star$, M$_{\star, \rm{bulge}}$, and $\Sigma_{5}$.}
    \label{fig:frac_bins}
\end{figure}

Similar results were found by \cite{oh+12} with a sample of  $\sim6600$ galaxies. However, these authors suggest that the comparison of total fractions might not be sufficient to reveal the entire range of bar effects. They show that the fraction of barred galaxies and AGN can be highly affected by galaxy properties (such as mass and morphology). 
For this reason, \cite{oh+12} divided their sample of galaxies in bins according to intervals of stellar mass. Their results suggest that bar effects on AGN fractions are only significant in intermediate-mass galaxies, in a range between 10$^{10.5}$ and 10$^{11}$~M$_{\odot}$. We performed the same test here, dividing our galaxies in intervals of stellar mass in our matched sample (by  M$_\star$, M$_{\star, \rm{bulge}}$, and $\Sigma_{5}$), as shown in Fig.~\ref{fig:frac_bins}. 
Our results differ slightly from those of \cite{oh+12}. 
This may be mainly due to the matching procedure. 
These latter authors found that AGN are more frequent in bluer and less massive galaxies. For this reason, it is important to define the samples of active and inactive galaxies in such a way that they have the same mass distributions before testing for fractions. Because we were careful in selecting the same mass distribution for active and inactive galaxies, the average fraction of AGN in each mass bin for the whole sample (when galaxies are not separated into barred and unbarred) is the same.
When we separate galaxies into barred and unbarred, the results from Fig.~\ref{fig:frac_bins} suggest that the fraction of AGN in barred galaxies is more significant for intermediate and more massive galaxies (similar to the findings of \citet{oh+12}). These results suggest that since gas is found more abundantly in the less massive galaxies, bars would not make a difference in helping to build a gas reservoir to feed the SMBH, by bringing gas to the central regions from larger radii.

\subsection{The relation between AGN activity level and the presence of a bar}

If bars are connected with AGN activity, one might expect to find correlations between the presence and properties of the bar and the activity level. Because we are not comparing active and inactive galaxies here, we use the whole AGN sample before the matching process. There are 94 AGN, of which 52 have bars and 42 do not. In addition, barred and unbarred AGN have $\Sigma_5$ distributions that do not indicate differences between the environments of these objects (AD-test with p-values > 0.25).

One of the best tracers of AGN activity in the optical is the $[\ion{O}{III}]\lambda$5007 emission line, which is not severely affected by stellar population contamination \citep{heckman+04}. Here we use the SDSS emission lines measured by \citet{Kauffmann2003}, corrected by extinction. In addition, \citet{oh+12} suggested that to show the difference in emission luminosity between barred and unbarred galaxies free from the mass–luminosity relation, it is better to use the specific emission luminosity, which is defined as the emission line luminosity divided by the fibre luminosity ($L[\ion{O}{III}] / L_{x, \rm fibre }$). We present the results from tests for both indicators below. 

\begin{figure}
    \centering
    \includegraphics[width=8cm]{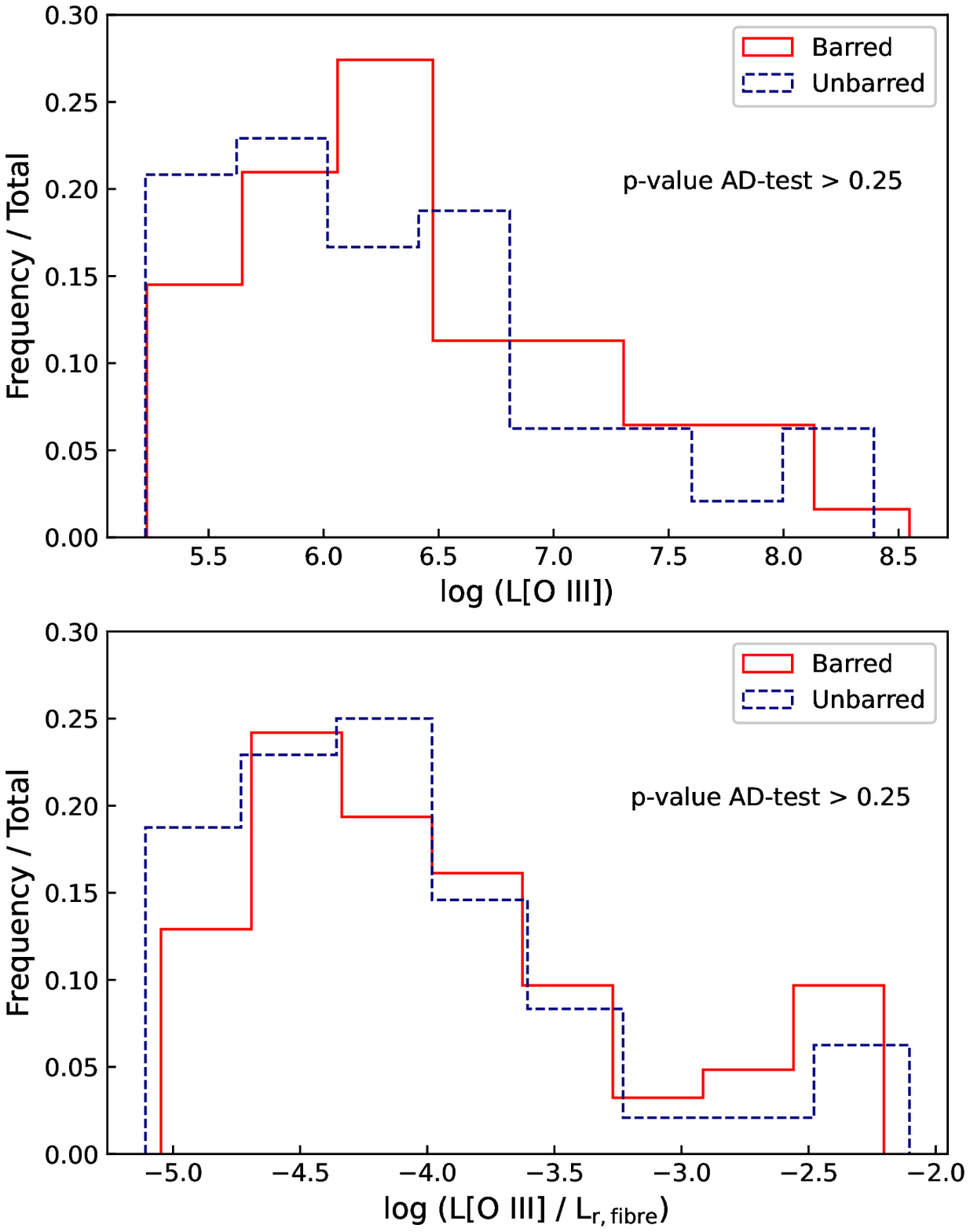}
    \caption{Distribution of $[\ion{O}{III}]$ luminosity. Top panel: Distribution of $[\ion{O}{III}]\lambda 5007$ luminosity for barred and unbarred galaxies. Bottom panel: Distribution of $L[\ion{O}{iii}]/L_{r, \rm fibre }$ for barred and unbarred galaxies. In both panels, the continuous red line represents the distribution for barred galaxies, while the dashed blue line is the distribution for unbarred galaxies. Neither test allows us to claim the existence of differences in the distributions of these parameters regarding morphology.}
    \label{fig:hist_oiii_hb_atv}
\end{figure}

The top panel of Fig.~\ref{fig:hist_oiii_hb_atv} shows the comparison between the $[\ion{O}{III}]\lambda5007$ luminosity for barred and unbarred active galaxies. The Anderson-Darling test shows a p-value of greater than 0.25, which is insufficient to claim a significant difference between the two distributions. In the bottom panel of the same figure it is shown that for $L[\ion{O}{III}] / L_{r, \rm fibre }$ we also cannot consider that there are differences in the underlying distributions of this parameter between barred and unbarred galaxies. However, \citet{alonso+18}, who made a similar comparison using only $L[\ion{O}{III}],$ found that barred AGN galaxies tend to have higher $[\ion{O}{III}]$ luminosities, with different distributions according to a KS-test, but they did not remove composites or weak LINERs from their sample. However, we cannot rule out that different results might be a consequence of very different sample sizes (these latter authors have a sample of $\sim1000$ barred galaxies with AGN).

A more direct test of the effect of the bar on the AGN feeding process could be performed through the accretion strength of the black hole. As a proxy of accretion, we employed the accretion rate parameter $\mathcal{R}$, defined by \citet{heckman+04} as the logarithm of the ratio between the $[\ion{O}{III}]$ luminosity and the black hole mass, in solar units.
The black hole mass was determined following \citet{graham+11}, who derives different M$-\sigma$ coefficients for unbarred and barred galaxies as described in Sect.~\ref{subsec:blackhole} using Eq.~\ref{eq:m-sigma_nao-barrada} and Eq.~\ref{eq:m-sigma_barrada}.
For our sample, we considered only galaxies with stellar velocity dispersion $\sigma_\star>70$~km~s$^{-1}$, because the SDSS spectrograph resolution is $\approx 60-70~$km~s$^{-1}$. This cut simply represents  an extra caution because only two galaxies in the sample were removed according to this criterion and the conclusions obtained were the same.

\begin{figure}
\centering
\includegraphics[width=8cm]{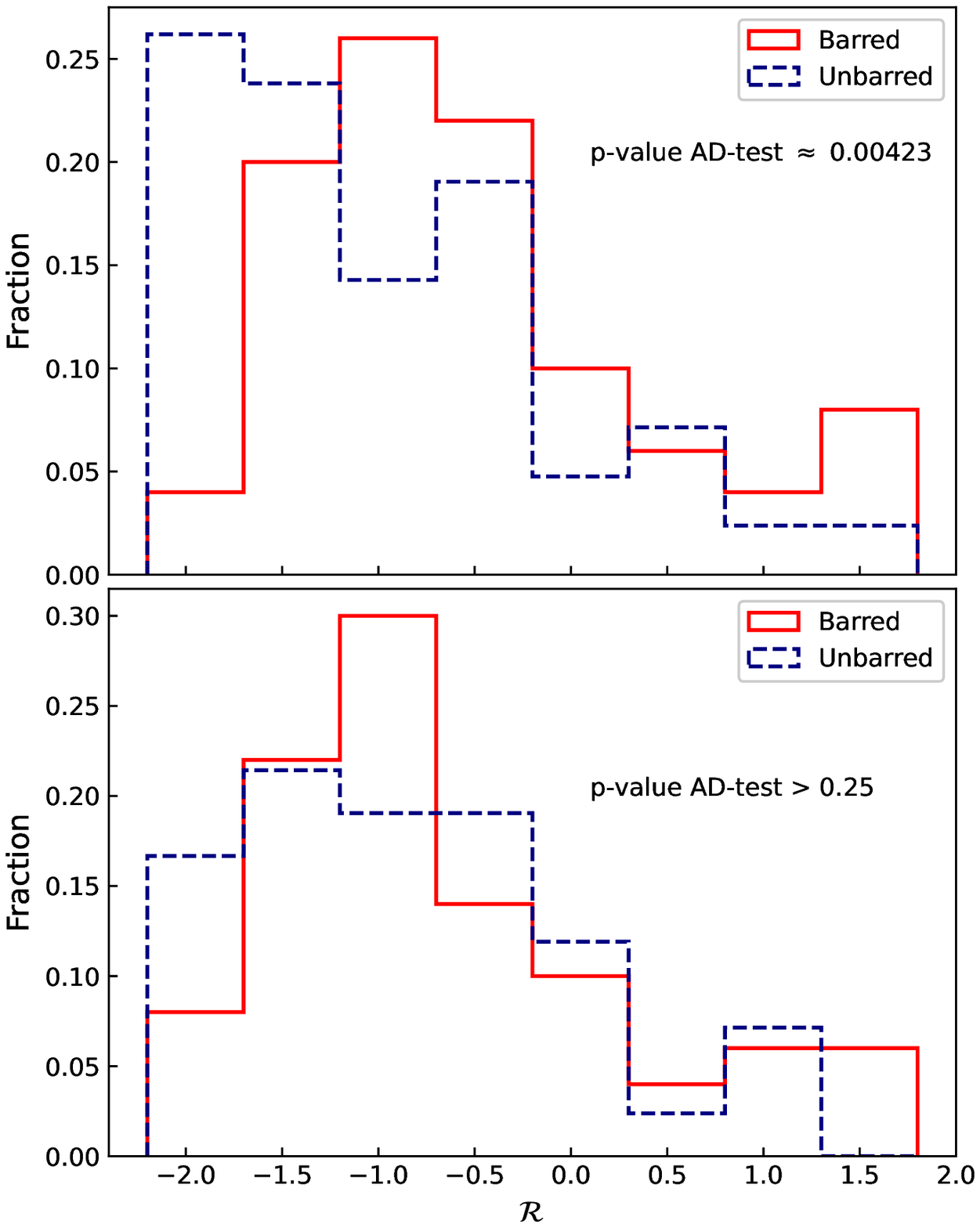}
\caption{Distribution of accretion parameter $\mathcal{R}$ for barred (solid red line) and unbarred (dotted blue line) AGN. In the top panel, the black hole masses were derived with different M--$\sigma$ relations that take into account the morphology as described in subsection \ref{subsec:blackhole} with Eq.~\ref{eq:m-sigma_barrada} and Eq.~\ref{eq:m-sigma_nao-barrada}. In the bottom panel, the black hole masses were derived with the M--$\sigma$ relation obtained for a sample without distinction regarding morphology (Eq.~\ref{eq:m-sigma_common}).}
\label{fig:R-param}
\end{figure}

 In the top panel of Fig.~\ref{fig:R-param} we present the distribution of $\mathcal{R}$ for barred and unbarred AGN. From this figure it is possible to see that barred galaxies tend to have a higher accretion parameter than unbarred galaxies. The Anderson-Darling test gives a p-value of 0.0042, meaning they are unlikely to have the same distribution. This result agrees with the findings of \citet{alonso+18}, who also used different equations to determine the black hole mass of barred and unbarred galaxies.

\citet{graham+11} also gives  coefficients for the M$-\sigma$ relation when barred and unbarred galaxies are considered together. The dispersion for this relation is larger, but if we use it to determine the black hole mass, the difference in the accretion parameter distribution almost disappears, as shown in the bottom panel of Fig.~\ref{fig:R-param}. \citet{alonso+18} performed the same test, but they found that even when using the equation derived for both barred and unbarred galaxies together, the distributions are still different, with barred galaxies having higher $\mathcal{R}$ values (although less pronounced). 
In our case, an Anderson-Darling test gives a p-value of greater than 0.25, which is not enough to claim that the distributions are different. Using a common M--$\sigma$ to estimate SMBH mass, \citet{galloway+15} also did not find enough significance in the KS-test to point out that the distributions are different in a sample with approximately $680$ AGN.
 
\subsection{The relation between AGN activity level and bar strength}

The approach of treating the influence of bars in galactic dynamics without considering its `strength' can be inaccurate \citep{cisternas+13}. The non-axisymmetric potential of a bar can have an enormous effect on the galaxy potential \citep{sellwilkin1993, martin95, binney+09} but this depends on the bar strength, the definitin of which is not straightforward. The strength of the bar can be defined such that it reflects its ability to apply torques to the ISM as a result of its gravitational potential. Attempts to quantify the influence of the bar have been made either directly through its properties or through its properties in relation to the properties of the host galaxy.

\begin{figure}
\centering
\includegraphics[width=0.95\linewidth]{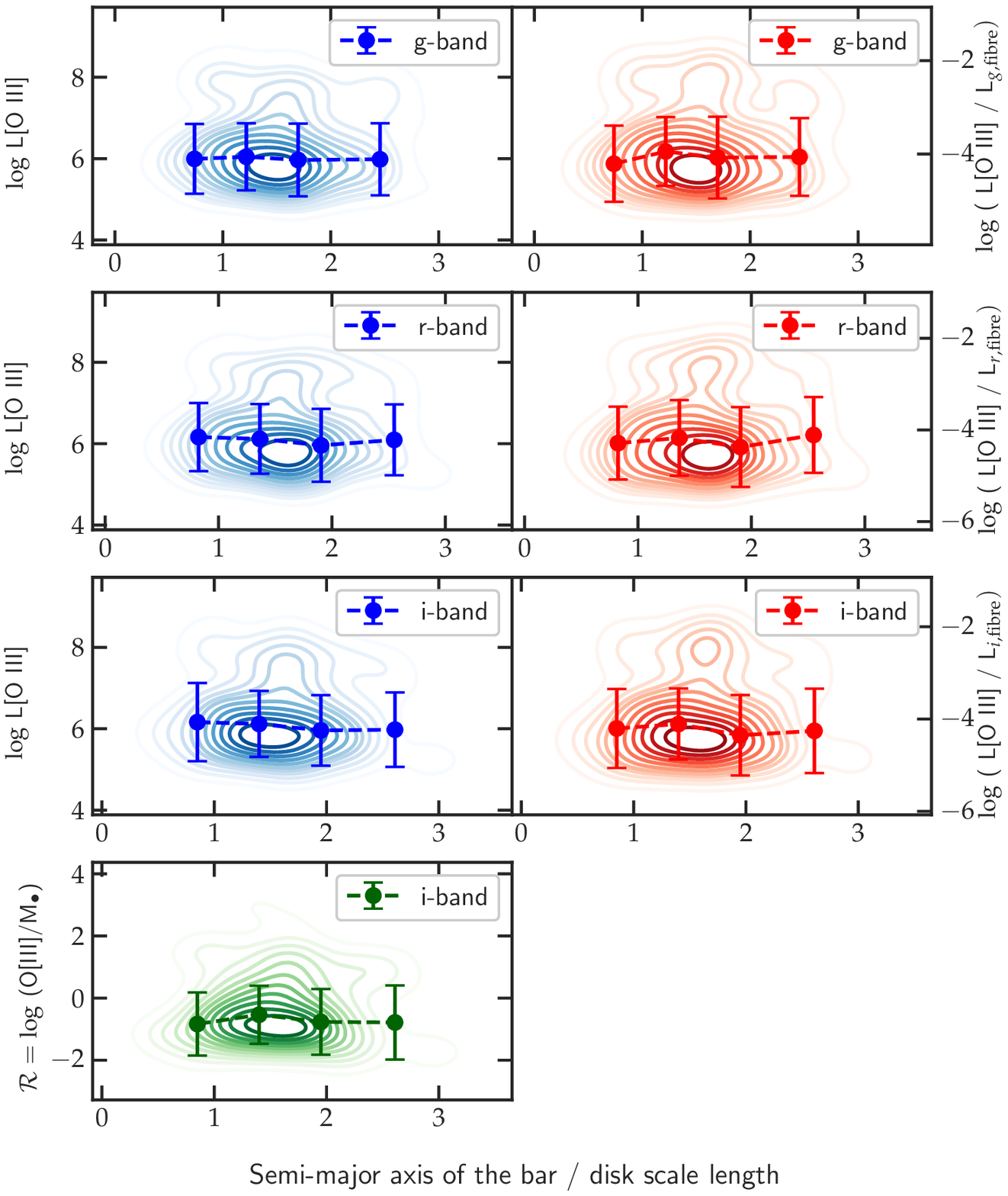}
\caption{$[\ion{O}{III}]\lambda 5007$ luminosity, $[\ion{O}{III}]\lambda 5007$ luminosity normalised by the fibre luminosity obtained for an aperture of 3$^{\prime \prime}$, and $\mathcal{R}$ plotted against the bar's semi-major axis normalised by the disk scale in $g$-,$r$-, and $i$-band.}
\label{fig:L-scalelength}
\end{figure}

\begin{figure}
\centering
\includegraphics[width=0.95\linewidth]{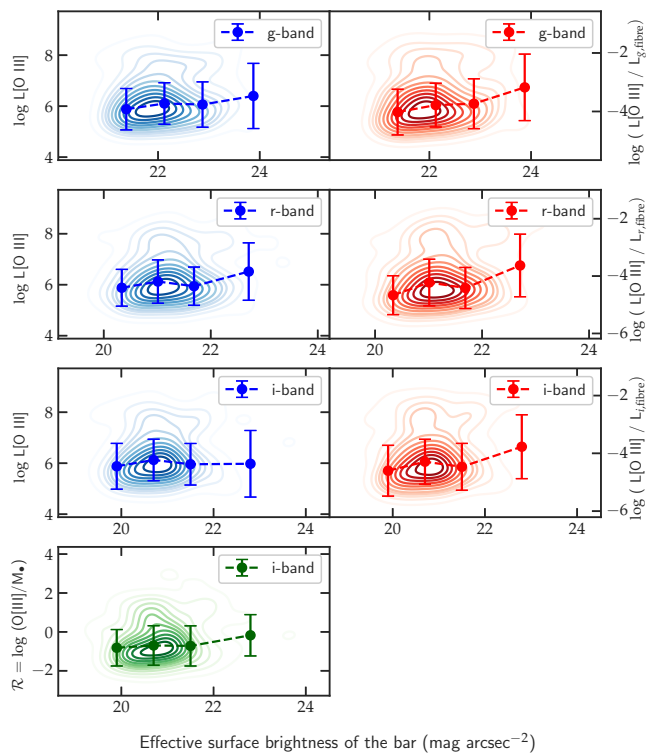}
\caption{$[\ion{O}{III}]\lambda 5007$ luminosity, $[\ion{O}{III}]\lambda 5007$ luminosity normalised by the fibre luminosity obtained for an aperture of 3$^{\prime \prime}$, and $\mathcal{R}$ plotted against the effective superficial brightness of the bar in the $g-$, $r-$, and $i$-band.}
\label{fig:surf_brig}
\end{figure}

\begin{figure}
\centering
\includegraphics[width=0.95\linewidth]{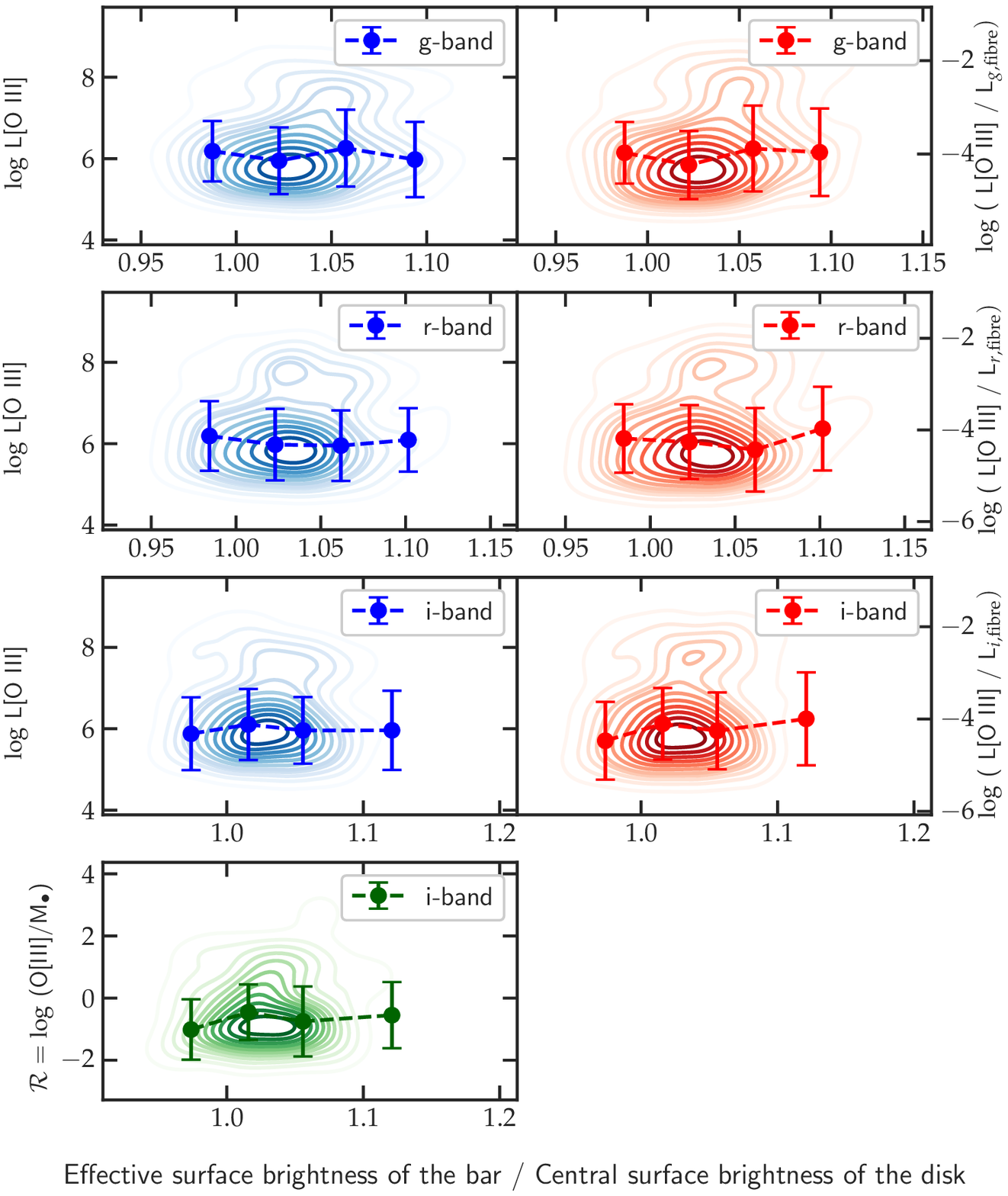}
\caption{$[\ion{O}{III}]\lambda 5007$ luminosity, $[\ion{O}{III}]\lambda 5007$ luminosity normalised by the fibre luminosity obtained for an aperture of 3$^{\prime \prime}$, and $\mathcal{R}$ plotted against the effective superficial brightness of the bar normalised by the central superficial brightness of the disc in the $g-$, $r-$, and $i$-band.}
\label{fig:surf_brig_norm}
\end{figure}

In order to investigate direct correlations between the bar and AGN activity, we now compare indicators of AGN activity with indicators of bar strength. As a bar strength indicator, we first use the size of the bar. For that we used the semi-major axis of the bar normalised by the disc scale length, measured in the three different SDSS bands. 
Another proxy of bar strength is the bar effective surface brightness in mag~arcsec$^{-2}$ for bands $g$, $r,$ and $i$ of SDSS.
We also normalised the bar effective surface brightness by the central surface brightness of the disc.
The comparison between these three bar strength indicators and AGN strength indicators $\log L{[\ion{O}{III}]}$, $\log (L_{[\ion{O}{III}]}/ L_{x,\rm{fibre }}$), and $\mathcal{R}$ can be seen in Figs.~\ref{fig:L-scalelength}, \ref{fig:surf_brig}, and \ref{fig:surf_brig_norm} respectively. 
No correlation was found between the AGN strength indicators and these indicators of bar strength. Most of the profiles are obviously flat. For the normalised bar effective surface brightness, there seems to be a tendency for higher activity at higher values, but given the error bars, no correlation can be claimed.
We employed the Spearman's rank correlation coefficient $\rho$ to quantify the correlations between these parameters, and results are shown in Table~\ref{tab:spearman_corr}. In this test, a coefficient $\rho$ close to 1 (or -1) indicates a correlation (or anti-correlation) between the parameters. We find no indication of correlations between AGN activity level and bar strength in any of the parameter combinations tested.

\begin{figure}
\centering
\includegraphics[width=0.8\linewidth]{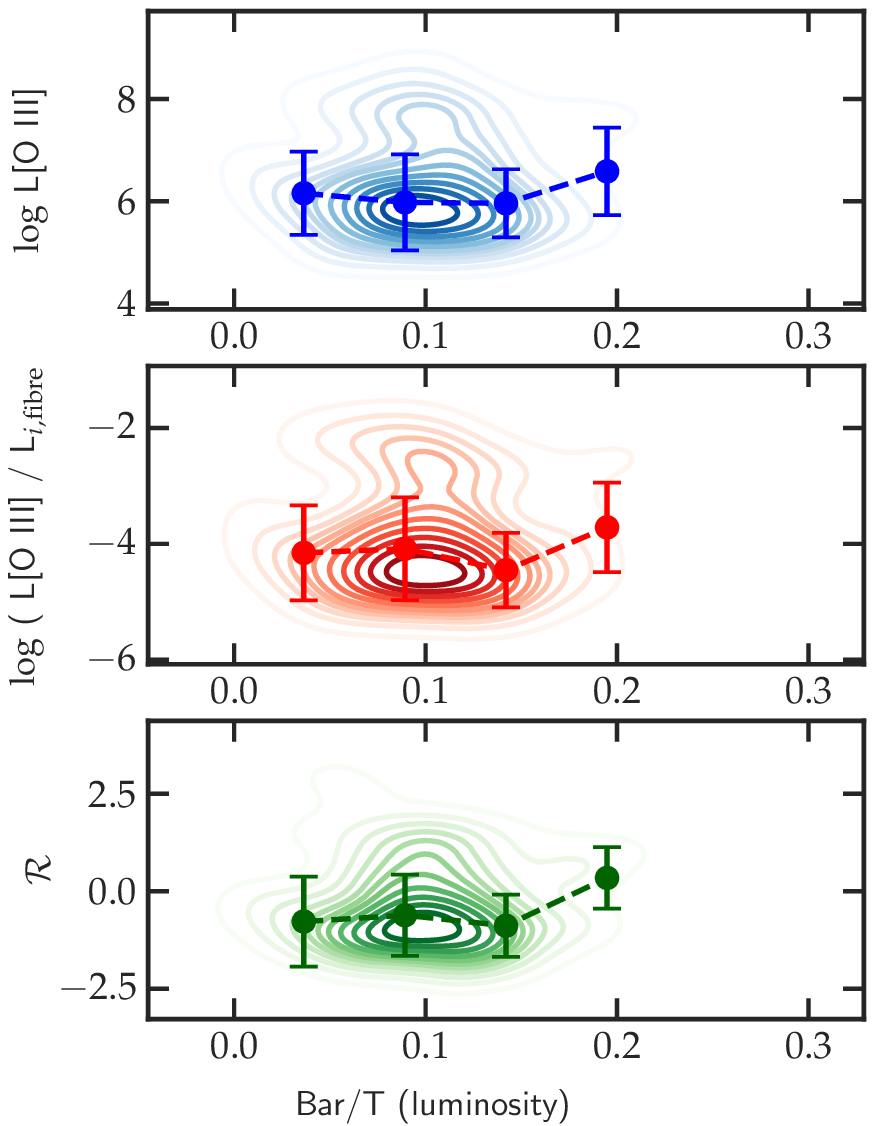}
\caption{$[\ion{O}{III}]\lambda 5007$ luminosity,  $[\ion{O}{III}]\lambda 5007$ luminosity normalised by the fibre luminosity, obtained for an aperture of 3$^{\prime \prime}$, and $\mathcal{R}$ plotted against the luminosity of the bar in relation to the total luminosity of the galaxy in $i$-band of the SDSS.}
\label{fig:bar-t_vs_oiii}
\end{figure}

We can also analyse the influence of the bar on the galactic dynamics given its relative importance in the total luminosity of the galaxy. Here we use the ratio between the luminosity of the bar and the total luminosity (Bar/T) in the $i$-band. As mentioned in \citet{martin95} and \citet{binney+09}, it is in the infrared (IR) and near-infrared (NIR) that the influence of the bar is best revealed. 
The $i$-band is the band most suited 
{to} our case because it is centred on 7480\AA. In addition to $\log L_{[\ion{O}{III}]}$ and normalised $\log L_{[\ion{O}{III}]}$, we compare Bar/T with the accretion parameter $\mathcal{R}$ \citep{heckman+04}. This is presented in Fig.~\ref{fig:bar-t_vs_oiii}, which indicates that the galaxies with the most prominent bars (last bin) tend to have higher activity levels on average, but again, given the scatter this result is not conclusive.
Moreover, the correlation tests performed do not show any correlation between these parameters. The results of Spearman's rank correlation coefficient $\rho$ and its significance for B/T and proxies of AGN activity are also shown in Table~\ref{tab:spearman_corr}. The coefficient for correlation between Bar/T and $\log L_{[\ion{O}{iii}]}$ is $-0.032$, whereas the same coefficient for Bar/T with respect to $\log ( L_{[\ion{O}{iii}]} / L_{i, \rm{fibre }})$ and $\mathcal{R}$ is $-0.041$ and $-0.005,$ respectively.

\begin{figure}
\centering
\includegraphics[width=\linewidth]{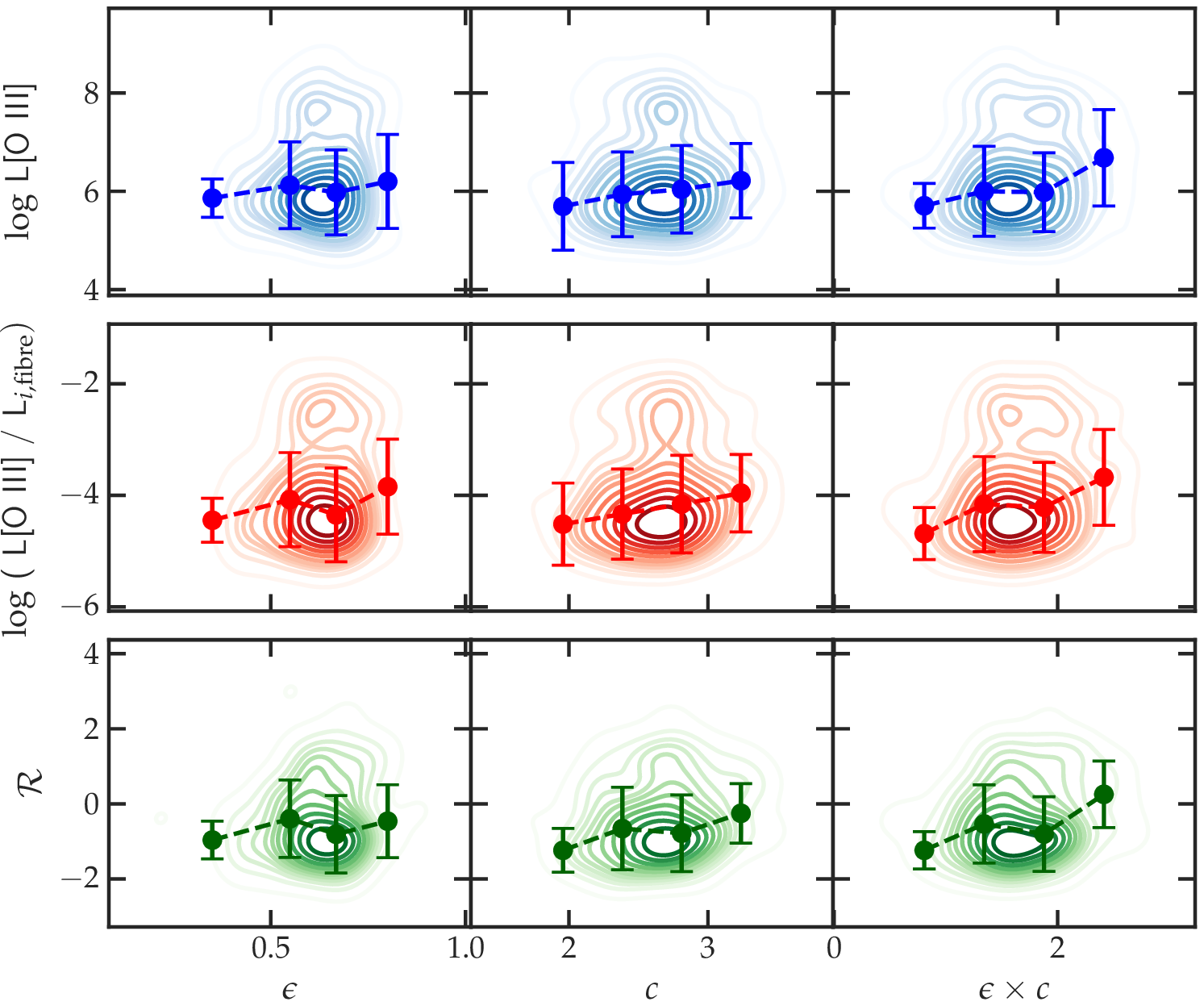}
\caption{Comparison of parameters related to bar strength (vertical axis) in relation to activity plotting parameters (horizontal axis). In the vertical axis, $\epsilon$ is the ellipticity of the bar and $c$ is the boxiness parameter, both measured in $i$-band of the SDSS. The horizontal axis shows the $[\ion{O}{III}]\lambda 5007$ luminosity,  $[\ion{O}{III}]\lambda 5007$ luminosity normalised by the fibre luminosity, obtained for an aperture of 3$^{\prime \prime}$, and $\mathcal{R}$.}
\label{fig:oiii_vs_strength}
\end{figure}

Bars show a range of axis ratios, from oval distortions to very elongated and well-defined structures. \citet{martin95} proposed that the axis ratio of the bar could be used as a quantification of bar strength, because it is directly connected to how non-axisymmetric the bar potential is. This ratio has the advantage of being obtained precisely and directly from photometric observations. 
Furthermore, it does not require any mass--luminosity assumption or other measurements. This means that the ellipticity ($\epsilon = 1 - b/a$) can be used to measure the bar strength.
In addition to ellipticity, the description of the bar's isophotes as generalised ellipses (Eq.~\ref{eq:fit_bar}) introduces the boxiness parameter $c$. There are indications in several studies that, during the secular evolution of the bar, while its pattern speed decreases, the bar becomes more elongated and also has more rectangular isophotes resulting in a more boxy morphology ($c > 2$) \citep{athanassoula2002, gadotti2011, cheung+13}.
Also there is evidence for a correlation between the boxiness parameter and ellipticity \citep{gadotti2011}. 
Given this correlation, \cite{gadotti2011} proposes the product $\epsilon \times c$ as a proxy for the bar 
strength. Here, we compare $\epsilon$, $c,$  and $\epsilon \times c$ against $\log L_{[\ion{O}{III}]}$, $\log (L_{[\ion{O}{III}]}/ L_{i,\rm{fibre }}),$ and $\mathcal{R}$ in Fig.~\ref{fig:oiii_vs_strength}. Correlation tests performed for these AGN activity level and bar strength indicators are presented in Table~\ref{tab:spearman_corr} and also indicate that there is no correlation.

\begin{table}
\caption{Spearman's Rank correlation coefficient $\rho$.}
\label{tab:spearman_corr}
\begin{tabular}{llccc}
\hline\hline
\multicolumn{1}{c}{Property} &   & Activity proxy                               & $\rho$ & Significance  \\ \hline
$a_{\text{bar}}$ / $h$           & vs & $\log L[\ion{O}{iii}]$                  & -0.020 & 0.839         \\
(g-band)                         &    & $\log (L[\ion{O}{iii}] / L_{g,\rm{fibre }})$  & -0.018 & 0.848    \\ \hline
$a_{\text{bar}}$ / $h$           & vs & $\log L[\ion{O}{iii}]$                  & -0.002 & 0.985         \\
(r-band)                         &    & $\log (L[\ion{O}{iii}] / L_{r,\rm{fibre }})$  & 0.015  & 0.880         \\ \hline
$a_{\text{bar}}$ / $h$           &    & $\log L[\ion{O}{iii}]$                  & -0.016 & 0.870         \\
(i-band)                         & vs & $\log (L[\ion{O}{iii}] / L_{i,\rm{fibre }})$  & 0.003  & 0.973    \\ 
                                 &    & $\mathcal{R}$                           & -0.005 & 0.956         \\ \hline
$\mu_{e,\text{bar}}$             & vs & $\log L[\ion{O}{iii}]$                  & 0.095  & 0.323         \\
(g-band)                         &    & $\log (L[\ion{O}{iii}] / L_{g,\rm{fibre }})$  & 0.142  & 0.142         \\ \hline
$\mu_{e,\text{bar}}$             & vs & $\log L[\ion{O}{iii}]$                  & 0.102  & 0.291         \\
(r-band)                         &    & $\log (L[\ion{O}{iii}] / L_{r,\rm{fibre }})$  & 0.138  & 0.153         \\ \hline
$\mu_{e,\text{bar}}$             & vs & $\log L[\ion{O}{iii}]$                  & 0.097  & 0.315         \\
(i-band)                         &    & $\log (L[\ion{O}{iii}] / L_{i,\rm{fibre }})$  & 0.127  & 0.189    \\
                                 &    & $\mathcal{R}$                           & 0.138 & 0.153          \\ \hline
$\mu_{e,\text{bar}}$ / $\mu_{0}$ & vs & $\log L[\ion{O}{iii}]$                  & 0.109  & 0.258         \\
(g-band)                         &    & $\log (L[\ion{O}{iii}] / L_{g,\rm{fibre }})$  & 0.170  & 0.078         \\ \hline
$\mu_{e,\text{bar}}$ / $\mu_{0}$ & vs & $\log L[\ion{O}{iii}]$                  & 0.057  & 0.554         \\
(r-band)                         &    & $\log (L[\ion{O}{iii}] / L_{r,\rm{fibre }})$  & 0.120  & 0.214         \\ \hline
$\mu_{e,\text{bar}}$ / $\mu_{0}$ &    & $\log L[\ion{O}{iii}]$                  & 0.011  & 0.910         \\
(i-band)                         & vs & $\log (L[\ion{O}{iii}] / L_{i,\rm{fibre }})$  & 0.070  & 0.469     \\ 
                                 &    & $\mathcal{R}$                           & 0.088  & 0.364         \\ \hline
Bar / T                          &    & $\log L[\ion{O}{iii}]$                  & -0.032 & 0.744         \\
(lum.)                           & vs & $\log (L[\ion{O}{iii}] / L_{i,\rm{fibre }})$  & -0.041 & 0.669         \\
                                 &    & $\mathcal{R}$                           & -0.005 & 0.959         \\ \hline
                                 &    & $\log L[\ion{O}{iii}]$                  & 0.051  & 0.601         \\
$\epsilon$                       & vs & $\log (L[\ion{O}{iii}] / L_{i,\rm{fibre }})$  & 0.076  & 0.433         \\
                                 &    & $\mathcal{R}$                           & 0.085  & 0.381         \\ \hline
                                 &    & $\log L[\ion{O}{iii}]$                  & 0.060  & 0.537         \\
$c$                              & vs & $\log (L[\ion{O}{iii}] / L_{i, \rm{fibre }})$ & 0.116  & 0.223         \\
                                 &    & $\mathcal{R}$                           & 0.152  & 0.114         \\ \hline
                                 &    & $\log L[\ion{O}{iii}]$                  & 0.074  & 0.445         \\
$\epsilon \times c$              & vs & $\log (L[\ion{O}{iii}] / L_{i,\rm{fibre }})$  & 0.118  & 0.222         \\
                                 &    & $\mathcal{R}$                           & 0.136  & 0.157         \\ \hline
\end{tabular}
\tablefoot{Correlation tests for bar properties and proxies of AGN activity are summarised. The following properties are used to estimate the strength of the bar: Semi-major axis of the bar ($a_{\text{bar}}$) normalised by the disc scale length ($h$), effective surface brightness of the bar ($\mu_{e,\text{bar}}$), effective surface brightness of the bar normalised by central surface brightness of the disc ($\mu_{0}$). All of these properties are shown in g-, r-, and i-band. The Bar/T (lum.) ratio, ellipticity ($\epsilon$), bar boxiness ($c$), and $\epsilon \times c$ are also shown in i-band. The bar properties are tested for correlation with AGN activity proxies, such as the $\log [\ion{O}{iii}]$ luminosity in solar units, $\log [\ion{O}{iii}]$ normalised by fibre  luminosity, and accretion parameter $\mathcal{R}$ \citep{heckman+04}. For each test, the correlation coefficient and the significance are shown.}
\end{table}

\section{Discussion}
In all the cases studied here, the fraction of AGN hosts is systematically larger in barred galaxies than in unbarred galaxies. Given the multiple factors that can degenerate the analysis where we investigate a possible relationship between the presence of bars and AGN, the increase in the AGN fraction in barred galaxies compared to unbarred galaxies does not show high levels of significance, ranging from 1.5$\sigma$ to 2.3$\sigma$. This is expected, particularly because of the differences in the timescales of these processes. Still, the fact that we systematically find an excess of AGN hosts in barred galaxies with methodologies applied in an attempt to reduce bias, taken together with all theoretical and observational limitations for the direct detection of AGN feeding by bars, leads us to conclude that the presence of a bar does to some extent favour the existence of an AGN. Bars appear to contribute to some degree in the process of angular momentum removal from the gas that will feed the SMBH. If bars do not directly feed SMBHs, they appear to at least  contribute to the formation of a reservoir of gas close enough to the black hole ($\lesssim$ 100~pc), where other physical processes start to dominate in the final funneling of the gas. We note that, as discussed above, works that investigate a possible role of bars in the AGN feeding process find conflicting results. Studies with this purpose use samples with a wide range of sizes, and some of these works have larger samples than the one used here. We therefore cannot rule out that some of the differences we find in this work, compared to previous ones, might be due to our smaller sample size.

On scales of a few hundred parsecs, the `bars within bars' scenario together with m=1 instabilities and nuclear warps may also be another important piece of the fuelling process \citep{shlosman+98, schinnerer+00, hunt+08, bittner+21}.
Recently, probing the kinematics and morphology of the gas inside the central kiloparsec of galaxies became feasible with ALMA observations. Evidence of AGN feeding was found for example in NGC~1566 \citep{combes+14} and NGC~613 \citep{audibert2019}.
On scales of a few tens of parsecs, other mechanisms such as viscous torques can drive massive gas clouds to the nucleus \citep[e.g.][]{combes03, jogee+06}. Simulations suggest that a series of dynamical instabilities on these scales are involved in the fuelling process \citep{hopkins+10, hopkins+12}. 
A better understanding of how these small-scale processes work and how they relate to large-scale bars can only be achieved with the systematic study of the nuclear regions of barred and unbarred AGN. 
Integral field unit observations of these systems may provide the means to answer some of these questions \citep{gadotti+19}.

On the other hand, our tests show that, in our active galaxies, the connection between the level of activity and the presence of a bar is less clear. While we find that the presence of a bar does not play a role in the $[\ion{O}{III}]$ emission line luminosity or in the normalised L$[\ion{O}{III}]$, the accretion rate in barred galaxies is statistically significantly higher than in unbarred galaxies (with an Anderson-Darling p-value of 0.0042; see top panel in Fig.~\ref{fig:R-param}). However, the latter relationship depends on the M--$\sigma$ relation employed. If a single M--$\sigma$ relation is employed for both barred and unbarred galaxies, we find no difference in the corresponding distributions of accretion rate. However, using the individual M--$\sigma$ relations derived by \citet{graham+11} separately for barred and unbarred galaxies, we find a significant difference. We argue that indeed the correct approach is to use independent M--$\sigma$ relations, and therefore the higher accretion rates found in barred galaxies is relevant. Furthermore, because of the orbital properties of stars in bars, for a fixed bulge mass, barred galaxies show on average higher velocity dispersion within the SDSS fibre radius \citep[particularly for low-mass bulges; see e.g.][]{gadotti_kauffmann+09}. Therefore, it is important that this difference is accounted for in the separate M--$\sigma$ relations.

Finally, we find no evidence that the strength of the bar, as measured with several proxies, plays a role in the level of activity in our active galaxies. This suggests that while stronger bars may build larger gas reservoirs to feed the SMBH (if enough gas is available in the main galaxy disc), the size of the reservoir does not play a role in the level of activity in a given AGN episode in the life of the galaxy. This is consistent with the fact that the gas mass necessary to feed AGN activity ($\sim$ 10$^5$ - 10$^6$~M$_{\odot}$) is small compared to the typical gas mass available in an average disc galaxy ($\sim$ 10$^9$~M$_{\odot}$).

\section{Conclusions}

        An AGN is the result of gas accretion by a SMBH located in the centre of a galaxy. To sustain the luminosity observed during the expected lifetime of an AGN, a relatively large quantity of matter has to be brought to the centre of the galaxy to feed it. In this sense, a mechanism should be responsible for removing angular momentum from the gas of the external regions so that it can move inwards. Galactic bars are frequently invoked as an important mechanism in this process. This bar--AGN   connection has been investigated by a number of authors but results have been controversial. Part of the conflict between empirical results can be explained by the different methods used to determine galaxy activity and identify and measure bar properties. On top of that, theory predicts different timescales for these phenomena and despite many simulations showing that the bar is indeed an efficient mechanism to produce a reservoir in the central region of galaxies, many of them also show that the gas becomes to some extent trapped in the inner Lindblad resonances. Secondary mechanisms would then be necessary to bring this gas to feed the AGN.  
    
       In this work, we investigated the role of bars on AGN feeding using a sample of galaxies based on SDSS DR2. The galaxy images were decomposed by \citetalias{gadotti+09} into discs, bulges, and bars using photometry in the $g$, $r,$ and $i$ bands with the code \texttt{BUDDA}. Our main conclusions can be summarised as follows:
        
        \begin{enumerate}
         \item  At first, we compared the fraction of AGN in barred and unbarred galaxies. We minimised sample biases using the propensity score sample of active and inactive galaxies with the same distribution of key parameters, namely M$_\star$, M$_{\star,\rm{bulge}}$, $\Sigma_{5}$, colour (g-r), and Sérsic index. We find that AGN are more frequently found in barred galaxies, which supports the idea that the presence of a bar favours the AGN feeding process.
        
        \item We also find that the accretion parameter $\mathcal{R} = \log([\ion{O}{III}]/\rm{M_{\bullet}})$  tends to be  greater in barred galaxies  than in their unbarred counterparts. This result is obtained when we derive the mass of the black hole using different and specific M--$\sigma$ relations for barred and unbarred galaxies. When we use a common M--$\sigma$ relation to both barred and unbarred galaxies, we do not observe significant differences in this parameter. However, we argue that the correct approach is to use separate  M--$\sigma$ relations, and therefore the higher accretion rates seen in barred galaxies is relevant.

        \item If bars are the main mechanism responsible for feeding an AGN, one might expect a correlation between bar strength and AGN activity level. We searched for this correlation using  the $[\ion{O}{III}]$ luminosity, the $[\ion{O}{III}]$ luminosity normalised by the luminosity in the SDSS fibre, and the accretion parameter $\mathcal{R}$ as AGN activity indicators. For the bar strength, we investigated correlations with the size of the bar (semi-major axis of the bar normalised by the disc scale length), its effective surface brightness for each SDSS band (normalised by the central brightness of the disc in the corresponding band), its luminosity (ratio between the luminosity of the bar and the total luminosity in $i$-band), and geometric parameters (ellipticity, boxiness, and their product). We find no correlation between any of the activity and bar strength indicators. This appears to be a clear indication that, although bars might indeed be important to bring gas to the central regions of disc galaxies, building a reservoir of gas to feed the AGN, secondary processes are necessary to accomplish this feeding process. For example, \citet{audibert2019} observed the flow of gas inside the inner Lindblad resonance ring of the bar to the central region using ALMA observations.  
      \end{enumerate}

     The process of AGN feeding is very complex, involving large- and small-scale phenomena. Systematic and multiwavelength studies with high spatial resolution on the central regions of large samples of barred and unbarred galaxies are necessary to shed light on these processes.

\begin{acknowledgements}
    We are grateful to the anonymous referee for the thoughtful comments that were very helpful in improving this paper. This work was supported by The São Paulo Research Foundation - FAPESP with the grant 2018/24967-1. 
    L.M. thanks CNPQ for financial support through grant  306359/2018-9.
    PC acknowledges support from Conselho Nacional de Desenvolvimento Cient\'ifico e Tecnol\'ogico (CNPq) under grant 310041/2018-0 and 
    from Funda\c{c}\~{a}o de Amparo \`{a} Pesquisa do Estado de S\~{a}o Paulo (FAPESP) process number 2018/05392-8.
    Funding for the SDSS and SDSS-II has been provided by the Alfred P. Sloan Foundation, the Participating Institutions, the National Science Foundation, the U.S. Department of Energy, the National Aeronautics and Space Administration, the Japanese Monbukagakusho, the Max Planck Society, and the Higher Education Funding Council for England. The SDSS Web Site is http://www.sdss.org/.
    The SDSS is managed by the Astrophysical Research Consortium for the Participating Institutions. The Participating Institutions are the American Museum of Natural History, Astrophysical Institute Potsdam, University of Basel, University of Cambridge, Case Western Reserve University, University of Chicago, Drexel University, Fermilab, the Institute for Advanced Study, the Japan Participation Group, Johns Hopkins University, the Joint Institute for Nuclear Astrophysics, the Kavli Institute for Particle Astrophysics and Cosmology, the Korean Scientist Group, the Chinese Academy of Sciences (LAMOST), Los Alamos National Laboratory, the Max-Planck-Institute for Astronomy (MPIA), the Max-Planck-Institute for Astrophysics (MPA), New Mexico State University, Ohio State University, University of Pittsburgh, University of Portsmouth, Princeton University, the United States Naval Observatory, and the University of Washington.
\end{acknowledgements}

%
\bibliographystyle{aa} 
\bibliography{biblio.bib} 
%

\end{document}